\documentclass[
aip,
pof
amsmath,
amssymb, 
preprint,
]{revtex4-1}

\usepackage{dcolumn}
\usepackage{bm}
\usepackage[utf8]{inputenc} 
\usepackage[T1]{fontenc}    
\usepackage{hyperref}       
\usepackage{url}            
\usepackage{booktabs}       
\usepackage{amsfonts}       
\usepackage{nicefrac}       
\usepackage{microtype}      
\usepackage{amsmath}
\usepackage{amssymb}
\usepackage{graphicx}
\usepackage{tikz}
\usepackage{xcolor}
\usepackage{mathptmx}
\usepackage{etoolbox}
\usepackage{overpic}
\graphicspath{{figures/}}
\DeclareGraphicsExtensions{.pdf,.eps,.png,.jpg,.jpeg,.bmp}

\makeatletter
\def\@email#1#2{%
 \endgroup
 \patchcmd{\titleblock@produce}
  {\frontmatter@RRAPformat}
  {\frontmatter@RRAPformat{\produce@RRAP{*#1\href{mailto:#2}{#2}}}\frontmatter@RRAPformat}
  {}{}
}%
\makeatother
\begin{document}

\title{Learning second-order TVD flux limiters using differentiable solvers}

\author{Chenyang Huang}
\affiliation{Department of Mechanical Engineering, University of Michigan, Ann Arbor, Michigan 48109, USA}

\author{Amal S.~Sebastian}
\affiliation{Department of Aerospace Engineering, University of Michigan, Ann Arbor, Michigan 48109, USA
}
\author{Venkatasubramanian Viswanathan}
 \email{venkvis@umich.edu}
\affiliation{Department of Mechanical Engineering, University of Michigan, Ann Arbor, Michigan 48109, USA}
\affiliation{Department of Aerospace Engineering, University of Michigan, Ann Arbor, Michigan 48109, USA
}

\begin{abstract}
This paper presents a data-driven framework for learning optimal second-order total variation diminishing (TVD) flux limiters via differentiable simulations. In our fully differentiable finite volume solvers, the limiter functions are replaced by neural networks. By representing the limiter as a pointwise convex linear combination of the Minmod and Superbee limiters, we enforce both second-order accuracy and TVD constraints at all stages of training. Our approach leverages gradient-based optimization through automatic differentiation, allowing a direct backpropagation of errors from numerical solutions to the limiter parameters. We demonstrate the effectiveness of this method on various hyperbolic conservation laws, including the linear advection equation, the Burgers' equation, and the one-dimensional Euler equations. Remarkably, a limiter trained solely on linear advection exhibits strong generalizability, surpassing the accuracy of most classical flux limiters across a range of problems with shocks and discontinuities. The learned flux limiters can be readily integrated into existing computational fluid dynamics codes, and the proposed methodology also offers a flexible pathway to systematically develop and optimize flux limiters for complex flow problems.
\end{abstract}

\maketitle

\section{Introduction}

Hyperbolic conservation laws describe the physics of phenomena in which information propagates with finite speed, such as the Euler equations for gas dynamics \cite{feistauer2003mathematical}. A key challenge in numerically solving hyperbolic conservation laws is preventing non-physical oscillations near discontinuities, as high-order numerical schemes---while improving accuracy in smooth regions---tend to yield oscillatory solutions near these discontinuities \citep{fedorenko1963application, woodward1984numerical}. This occurs because of dispersion errors and the high-frequency content associated with discontinuities \cite{moukalled2016finite}. Such oscillations may be benign for linear problems but can lead to instabilities and nonphysical states in practical, nonlinear systems \citep{liska2003comparison}.

High-resolution schemes incorporating the boundedness property have been developed to address this issue \cite{harten1983high, osher1984high, jasak1999high}. These schemes are able to offer sharp resolution around steep-gradient regions without introducing spurious oscillations, while simultaneously preserve at least second-order accuracy in smooth regions. Past efforts include the hybrid method \citep{harten1972self}, the flux-corrected transport (FCT) method \citep{boris1973flux}, the total variation diminishing (TVD) schemes \citep{van1977towards4, harten1983high, harten1984class}, and the (weighted) essentially non-oscillatory ((W)ENO) schemes \citep{harten1987uniformly, jiang1996efficient}. Among these approaches, TVD schemes, originally developed by \citet{harten1983high} from a purely algebraic point of view, are a group of the most popular high-resolution schemes for solving hyperbolic conservation laws in practical applications \citep{denner2015tvd, shiea2020novel}. In comparison to TVD schemes, (W)ENO schemes offer superior preservation of local extrema and high-frequency details; however, in some cases their reliance on larger finite-difference stencils poses significant challenges for boundary conditions and unstructured mesh configurations \citep{zhang2016eno}. \citet{sweby1984high} systematized the study of TVD limiters through the flux formulation and derived what is now known as the Sweby diagram. In general, TVD flux limiters are used to adaptively modify the numerical scheme, allowing for high-order accuracy in smooth regions while reducing to lower-order, more diffusive schemes near discontinuities to avoid oscillations. They exhibit multiple desirable features such as monotonicity-preserving property, computational simplicity and efficiency, and at least second-order accuracy in smooth regions \citep{roe1985some, goodman1988geometric, arora1997well, leveque2002finite, vcada2009compact}. Flux limiters are a key tool in CFD to enhance the accuracy, stability, and efficiency of numerical simulations, particularly in situations involving complex flow phenomena with shocks and steep gradients.

Commonly used traditional limiter functions are classified into two categories: piecewise-linear functions (e.g., Minmod, Superbee \citep{roe1986characteristic}, monotonized central (MC)\citep{van1977towards3}) and global smooth nonlinear functions (e.g., van Leer \citep{van1974towards}, the van Albada family \citep{van1982comparative}, OSPRE \citep{waterson1995unified}). Over the past several decades, researchers have developed a large number of new flux limiters. \citet{kemm2011comparative} proposed modifications to several classical flux limiters for better accuracy, convergence, and reconstruction of the local extrema. For example, he introduced a new parameter to adapt the Superbee limiter to the third-order scheme. \citet{zhang2015review} presented a comprehensive review of existing TVD schemes and proposed a new CFL-independent flux limiter for steady-state calculations. \citet{tang2020construction} constructed three symmetric limiter functions based on the classical van Albada, van Leer, and PR-$\kappa$ limiters. All these flux limiters have a fixed mathematical formulation (occasionally with hyperparameters), so their performance is limited by the inherent assumptions and simplifications in their design, which may not always be optimal for all flow conditions. More recently, researchers have begun exploring data-driven methods and machine learning algorithms to develop better numerical approximations tailored to specific flow problems. Although some approaches aim to learn an entire numerical scheme via neural networks \citep{mishra2018machine, bar2019learning, morand2024deep}, these methods must be re-trained for each new problem and cannot be directly incorporated into production CFD codes. On the other hand, some data-driven research focuses on directly learning the flux limiter. \citet{lochab2021improved} optimized the limiter functions with the aid of fuzzy logic operators, but the order of accuracy and the TVD property are not guaranteed. \citet{nguyen2022machine} constructed the limiter function as a piecewise linear function and solves the coefficient by least squares regression based on the one-dimensional Burgers equation. However, the learned flux limiter is first-order, which makes it very diffusive. In addition, the generalizability of the flux limiter to other problems is not tested. \citet{schwarz2023reinforcement} developed a slope limiter that is independent of a empirical global parameter while providing an optimal slope in a second-order finite volume solver by leveraging deep learning and reinforcement learning techniques. However, small wiggles in their solutions for shock tube problems indicate the non-TVD nature of their algorithm. 

While these developments illustrate the promise of data-driven approaches to learn flux limiters, the insufficient integration of physical constraints during training not only complicates the enforcement of high-order accuracy and the TVD property but also compromises the generalization ability. In contrast, by making the physics solver itself differentiable, constraints such as conservation laws, stability requirements, and order of accuracy can be directly enforced in the training process. This avoids the common pitfall where learned models violate fundamental physical principles and can not generalize well. \emph{Differentiable physics} \citep{liang2020differentiable, ramsundar2021differentiable, newbury2024review} is a concept that has recently gained prominence within numerous scientific domains (e.g., learning and control \citep{de2018end}, quantum chemistry \citep{kasim2022dqc}, molecular dynamics \citep{greener2024differentiable}, tokamak transport \citep{citrin2024torax}). It is a paradigm that enables hybrid methods that unify machine learning with traditional numerical solvers by leveraging the power of automatic differentiation (AD) \citep{paszke2017automatic} or adjoint methods \citep{giles2000introduction}. Notably, several differentiable CFD solvers have emerged in this framework, including PhiFlow \citep{holl2020phiflow}, JAX-CFD \citep{Kochkov2021-ML-CFD}, and JAX-Fluids \citep{bezgin2023jax}. These tools facilitate backpropagation through the entire solver, making it possible to perform end-to-end gradient-based optimization that respects physical laws. For instance, \citet{bezgin2023jax} illustrated the use of differentiable solvers to perform end-to-end optimization of the numerical viscosity in the Rusanov flux.

In this paper, we propose a framework to learn the optimal second-order TVD flux limiter using differentiable finite volume (FV) solvers where we replace the limiter function by a neural network. The learned flux limiters can be quickly embedded into existing CFD code bases once trained and we demonstrate its intergration into OpenFOAM. The proposed framework is demonstrated on three representative hyperbolic conservation laws: linear advection, Burgers’ equation, and the Euler equations. To enforce the second-order TVD constraint, we represent the flux limiter as a pointwise convex linear combination of Minmod and Superbee. By virtue of AD, we can backpropagate all the way back to the parameters of the neural network and update them, despite the loss function being a complex composite function of the network parameters due to the iterative update process of the numerical solver.  More generally, this approach opens the doors for a family of numerical schemes that can  be trained using differentiable physics.

\section{Method}
In this section, we briefly introduce the finite volume (FV) schemes for solving three typical one-dimensional hyperbolic conservation laws, explain the concept of flux limiter via linear advection equation, illustrate how limiter functions are parametrized via multilayer perceptrons (MLPs), and conclude with a sketch of the differentiable physics framework for learning the second-order TVD flux limiter.

\subsection{Finite volume method}
Consider the numerical solution of the one-dimensional systems of hyperbolic conservation law:
\begin{equation}\label{eqn:hcl}
    \frac{\partial q}{\partial t} + \frac{\partial f(q)}{\partial x} = 0,
\end{equation}
where $q = q(x,t)$ denotes the state and $f(q)$ is the flux function. We discretize the spatial domain into $N$ uniform cells, where $[x_{i-1/2}, x_{i+1/2}]$ denotes the $i$th cell for $i \in \{1, 2, \dots, N\}$. Let $\Delta t$ be the time step and $\Delta x$ be the cell size. We define the cell average of the state inside the cell $i$ at time $t_n$ as:
\begin{equation}\label{eqn:cell-ave}
    Q_i^n = \frac{1}{\Delta x} \int_{x_{i-1/2}}^{x_{i+1/2}} q(x, t_n) dx.
\end{equation}
Integrating Eq.~(\ref{eqn:hcl}) over cell $i$ gives
\begin{equation}\label{eqn:spatial-integration}
\frac{d}{dt}\int_{x_{i-1/2}}^{x_{i+1/2}} q(x, t) dx = f(q(x_{i-1/2}, t)) - f(q(x_{i+1/2}, t)).
\end{equation}
Using the definition in Eq.~(\ref{eqn:cell-ave}), integrating Eq.~(\ref{eqn:spatial-integration}) from $t_n$ to $t_{n+1}$ and dividing by $\Delta x$ yields
\begin{equation}
    Q_{i}^{n+1} - Q_{i}^{n} = \frac{1}{\Delta x}\left[\int_{t_n}^{t_{n+1}}f(q(x_{i-1/2}, t))dt - \int_{t_n}^{t_{n+1}}f(q(x_{i+1/2}, t))dt \right].
\end{equation}
In this paper, we use the numerical schemes of the form
\begin{equation}\label{eqn:state-update}
    Q_{i}^{n+1} = Q_{i}^{n} - \frac{\Delta t}{\Delta x}(F_{i+1/2}^{n} - F_{i-1/2}^{n}),
\end{equation}
where $F_{i-1/2}^{n} \approx \frac{1}{\Delta t}\int_{t_n}^{t_{n+1}}f(q(x_{i-1/2}, t))dt$ is some numerical approximation to the average flux along $x = x_{i-1/2}$.

In the next section, we describe the specific formulations of the numerical fluxes $F_{i-1/2}$ for the linear advection problem. The detailed numerical schemes used for solving the Burgers' equation and Euler equations are documented in Appendix \ref{app:fvm}.

\subsection{Flux limiter}
We explain the concept of flux limiter using the linear advection equation
\begin{equation}\label{eqn:lin-adv}
    \frac{\partial q}{\partial t} + \frac{\partial (aq)}{\partial x} = 0,
\end{equation}
where the advection velocity $a > 0$. The numerical fluxes for the first-order upwind (FOU) and the Lax--Wendroff (LW) scheme read
\begin{equation}\label{eqn:linear-fou-lw}
    \begin{aligned}
        F_{i-1/2}^{\text{FOU}} &= aQ_{i-1}, \\
        F_{i-1/2}^{\text{LW}} &= aQ_{i-1} + \frac{a}{2}(1-\nu)(Q_{i}-Q_{i-1}),
    \end{aligned}
\end{equation}
where we drop the superscript of time step for convenience. Due to the stability requirement, the Courant--Friedrichs--Lewy (CFL) number, $\nu = a\Delta t/\Delta x$, is limited to the range of $(0,1]$. Note that the Lax--Wendroff scheme essentially adds an additional second-order correction term (also referred to as an antidiffusive flux) to the first-order upwind flux.  This treatment is readily generalizable to nonlinear scalar equations and systems of conservation laws, as shown in Appendix \ref{app:fvm}.

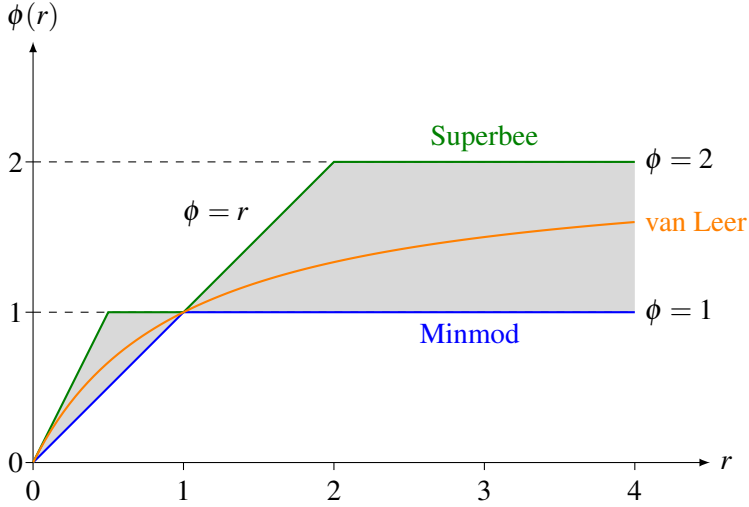
\begin{figure}[htbp]
    \centering
    \begin{tikzpicture}[>=latex,scale=2]
    
      \draw[->] (0,0) -- (4.5,0) node[right] {$r$};
      \draw[->] (0,0) -- (0,2.8) node[above] {$\phi(r)$};
    
      \fill[gray!30]
        (0,0) -- (0.5,1) -- (1,1) -- (2,2) -- (4,2)
        -- (4,1) -- (1,1) -- (0,0) -- cycle;

      \draw[dashed] (1,1) -- (2,2) node[midway,above left] {$\phi = r$};
      \draw[dashed] (0,1) -- (4,1) node[right] {$\phi=1$};
      \draw[dashed] (0,2) -- (4,2) node[right] {$\phi=2$};
    
      \draw[blue, thick] (0,0) -- (1,1) -- (4,1)
        node[midway,below right] {Minmod};
    
      \definecolor{ao(english)}{rgb}{0.0, 0.5, 0.0}
      \draw[ao(english), thick] (0,0) -- (0.5,1) -- (1,1) -- (2,2) -- (4,2)
        node[midway,above] {Superbee};

      \draw[orange, thick,domain=0:4,samples=100] 
        plot(\x,{2*\x/(1+\x)}) 
        node[right] {van Leer};
    
      \foreach \x in {0,1,2,3,4}
        \draw (\x,0) -- (\x,-0.05) node[below] {\x};
    
      \foreach \y in {0,1,2}
        \draw (-0.05,\y) -- (0,\y) node[left] {\y};
    
    \end{tikzpicture}
    \caption{A sketch of the second-order TVD region (shaded area), together with the classical flux limiters Minmod, Superbee, and van Leer.}
    \label{fig:second-order-tvd-region}
\end{figure}

We then introduce a flux limiter $\phi_{i-1/2}$, which linearly combines these two fluxes and gives the modified flux as
\begin{equation}\label{eqn:mod-flux}
\begin{aligned}
    F_{i-1/2} &= (1-\phi_{i-1/2})F_{i-1/2}^{\text{FOU}} + \phi_{i-1/2}F_{i-1/2}^{\text{LW}} \\
    &= aQ_{i-1} + \frac{a}{2}(1-\nu)\phi_{i-1/2}(Q_{i}-Q_{i-1}).
\end{aligned}
\end{equation} 
If $\phi_{i-1/2} = 1$ at all interfaces, which indicates local smoothness, we recover the non-TVD Lax--Wendroff scheme. On the other hand, if $\phi_{i-1/2} = 0$, we recover the TVD (but inaccurate) first-order upwind scheme. The value of $\phi_{i-1/2}$ is fully determined by the ratio of adjacent difference (slopes),
\begin{equation}\label{eqn:r-linear}
    r_{i-1/2} = \frac{\Delta Q_{i-3/2}}{\Delta Q_{i-1/2}} = \frac{Q_{i-1}-Q_{i-2}}{Q_{i}-Q_{i-1}}.
\end{equation}
The goal is to choose the limiter, $\phi(r)$, such that $\phi(1+\varepsilon) = 1 + \mathcal{O}(\varepsilon)$ in smooth regions for second-order accuracy \citep{roe1986characteristic}, while satisfying the TVD condition. According to Harten's lemma \citep{harten1983high}, a three-point stencil is TVD if
\begin{equation}
\begin{aligned}
    &\phi(r) = 0 \text{ for } r \leqslant 0, \\
    &\phi(r) \leqslant \min\left(\frac{2}{1-\nu},\frac{2r}{\nu}\right) \text{ for } r > 0.
\end{aligned}
\end{equation}
This condition is CFL number dependent. However, when applying the flux limiter to find steady solutions as the large-time limit of a (pseudo) unsteady flow, retaining the dependence on the CFL number shows little advantage. On the other hand, this condition has been empirically found to be too compressive for nonlinear systems and hence it is also better to use the more restrictive TVD constraints \cite{arora1997well}. These two scenarios lead to the simplified condition
\begin{equation}\label{eqn:tvd-condition}
    \phi(r) \leqslant \min(2,2r) \text{ for } r > 0.
\end{equation}
By taking a convex linear combination of the second-order Lax--Wendroff ($\phi(r) = 1$) and Beam--Warming ($\phi(r) = r$) schemes, \citet{sweby1984high} delineated a second-order TVD region that requires the flux limiter to satisfy Eq.~(\ref{eqn:tvd-condition}) and $\phi(1)=1$, which is illustrated in Fig.~\ref{fig:second-order-tvd-region}.

Additionally, a limiter is termed symmetric if it fulfills the condtion
\begin{equation}\label{eqn:sym-property}
    \frac{\phi(r)}{r} = \phi\left(\frac{1}{r}\right),
\end{equation}
thereby ensuring equivalent treatment of the top and bottom corners of a discontinuity.

\subsection{Parametrize second-order TVD flux limiters}
To enforce the second-order TVD constraint, we represent the flux limiter as a pointwise convex linear combination of Minmod and Superbee, i.e.,
\begin{equation}
    \phi_{\theta}(r) = (1 - \lambda_{\theta}(r))\phi_{\text{Minmod}}(r) + \lambda_{\theta}(r)\phi_{\text{Superbee}}(r),
\end{equation}
where
\begin{equation}
    \lambda_{\theta}(r) = \texttt{sigmoid}(g_{\theta}(r)).
\end{equation}
Here, $g_{\theta}(r)$ is parametrized by an MLP with learnable parameters $\theta$. Note that the linear combination needs to be convex so that $\phi_{\theta}(r)$ can lie in the second-order TVD region, which means that
\begin{equation}
    \lambda_{\theta}(r) \in [0,1], \quad \forall r.
\end{equation}
Thus, we apply the \texttt{sigmoid} activation function to the output of the MLP to get $\lambda_{\theta}(r)$. The MLP takes the slope ratio $r$ as input, has several hidden layers with the same number of neurons, and produces a scalar output $g_{\theta}(r)$. Each hidden layer $\ell$ applies a linear transformation followed by a pointwise nonlinear activation function $\sigma$. Concretely, if $h^{(0)} \in \mathbb{R}$ denotes the input, then the output of layer $\ell$ can be written as:
\begin{equation}
    h^{(\ell)} = \sigma\left(W^{(\ell)} h^{(\ell - 1)} + b^{(\ell)}\right),
\end{equation}
where $W^{(\ell)} \in \mathbb{R}^{m_\ell \times m_{\ell-1}}$ is the weight matrix, $b^{(\ell)} \in \mathbb{R}^{m_\ell}$ is the bias vector, and $m_\ell$ denotes the number of neurons in layer $\ell$. The activation function $\sigma$ can be chosen as \texttt{ReLU}, \texttt{tanh}, etc.

Directly parameterizing the limiter function by an MLP can lead to several issues. For instance, the learned flux limiter might not precisely pass through the point $(1,1)$, which would compromise accuracy even if it is very close. By contrast, the parameterization introduced here enforces constraints such as $\phi(r)=0$ for $r \leqslant 0$ and $\phi(1) = 1$. As a result, the parameter space can accommodate any limiter function lying within the second-order TVD region while guaranteeing that these critical properties are satisfied.

For convenience, this entire module is denoted as the neural flux limiter $f_{\theta}$, which takes $r$ as input and outputs the limiter function value $\phi_{\theta}(r)$.

\subsection{Differentiable simulations}\label{sec:diff-sim}
Differentiable simulations refer to simulations that provide not only the forward evolution of a system’s state over time but also accurate gradient information of that evolution with respect to model parameters or inputs \citep{liang2020differentiable, thuerey2021physics}. They offer a critical advantage over black-box machine learning models by directly integrating domain knowledge and numerical physics into the training process. By embedding AD into the solver’s numerical operations, differentiable simulations ensure that every operation can be accurately differentiated, thereby facilitating gradient-based learning and optimization.

\begin{figure}[htbp]
    \centering
    \includegraphics[width=\linewidth]{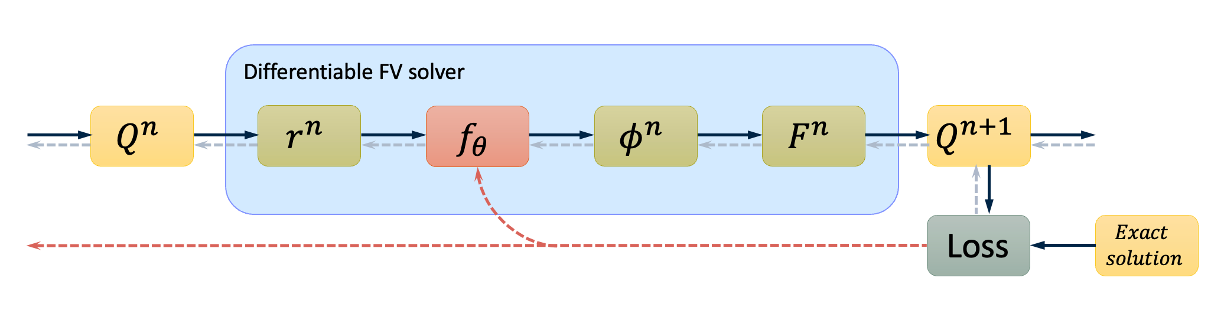}
    \caption{A schematic diagram of the differentiable simulation for learning second-order TVD flux limiters. The black solid arrows represent the forward pass to calculate the loss while the gray and red dashed arrows represent the backward pass to compute the gradients to update the parameters of the neural network $f_{\theta}$. At each time step, the smoothness measure $r^{n}$ at cell interfaces is calculated using the current states $Q^{n}$. The neural network $f_{\theta}$ takes $r^{n}$ as input and outputs the value $\phi^{n}$, which is used to evaluate the second-order correction term added to the underlying first-order flux. The states $Q^{n+1}$ at the next time step is then updated by the total flux $F^{n}$. The solution trajectory is propagated until prescribed time steps and the loss is evaluated with respect to the exact solution. The loss can be backpropagated through time using automatic differentiation to update the parameters of the neural flux limiter $f_{\theta}$ as all parts of the solution algorithm are fully differentiable.}
    \label{fig:schematic-diagram}
\end{figure}

A fully differentiable simulation pipeline to learn a second-order TVD flux limiter is shown in Fig.~\ref{fig:schematic-diagram}. At time step $n$, a smoothness measure $r^{n}$ is computed from the current states $Q^{n}$ based on the FV schemes. Then $r^{n}$ is fed into a neural network $f_{\theta}$ that gives us the value of $\phi^{n}$. This is used to form a second-order correction term added to the underlying first-order flux, yielding an updated flux $F^{n}$ that advances the states to $Q^{n+1}$. The solution is propagated for a prescribed number of time steps, and the loss is minimized via backpropagation through time using gradients obtained from AD.

\section{Datasets and training}
\subsection{Datasets}
\subsubsection{Linear advection}
The training dataset for linear advection is generated using the code from PDEBench \cite{takamoto2022pdebench}. We generate $10000$ trajectories with the periodic boundary condition and split them into $8192$ trajectories for the training set, $1024$ trajectories for the validation set, and $784$ trajectories for the test set. In our dataset, the initial condition is the superposition of two sinusoidal waves
\begin{eqnarray}
    q_0(x) = \sum_{i=1}^{2}A_i\sin(k_i x + \varphi_i),
\end{eqnarray}
where $k_i = 2\pi n_i/L_x$ are wave numbers with $n_i \in \{1, 2, \dots,8\}$, $L_x = 1$ is the domain size. The amplitude $A_i$ and phase $\varphi_i$ are uniformly sampled from $[0,1]$ and $(0, 2\pi)$, respectively. In addition, we randomly apply the absolute value function and the window-function to increase the complexity of the dataset.

As stated in \cite{takamoto2022pdebench}, the numerical solutions are computed using the temporally and spatially second-order upwind finite difference scheme. The spatial domain $[0,1]$ is discretized to $1024$ cells with $\Delta x = 1/1024$. The advection velocity is $a = 1.0$, and the CFL number is set as $\nu = 0.4$, from which we know the time step $\Delta t = \nu\Delta x$. We advect all of the initial conditions for $0.125$ s, which is $320$ time steps. Data is saved every time step.

\subsubsection{Burgers' equation}
The initial conditions are identical to those in the linear advection dataset. We solve the Burgers’ equation by setting the diffusion coefficient as $3\times 10^{-4}$ to enhance solver stability, using a CFL number of $\nu=0.4$. The numerical solver employs the temporally and spatially second-order upwind difference scheme for the nonlinear advection term, and the central difference scheme for the diffusion term \citep{takamoto2022pdebench}. Data is saved every 0.01 seconds for 0.2 seconds, during which shocks are generated. 

\subsubsection{Euler equations}\label{sec:euler-dataset}
\begin{figure*}[htp]
    \centering
    \begin{tabular}{rr}
    \begin{overpic}[width=0.48\textwidth]{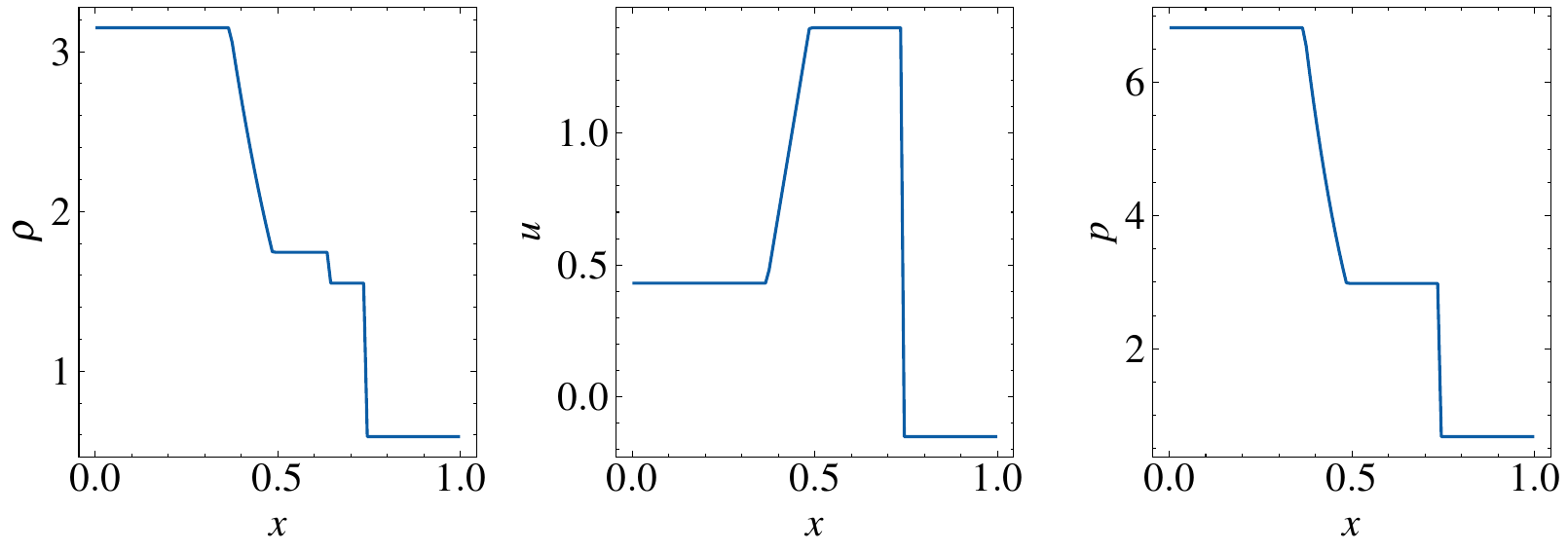}
        \put(-3,36){(a)}
    \end{overpic} &
    \begin{overpic}[width=0.48\textwidth]{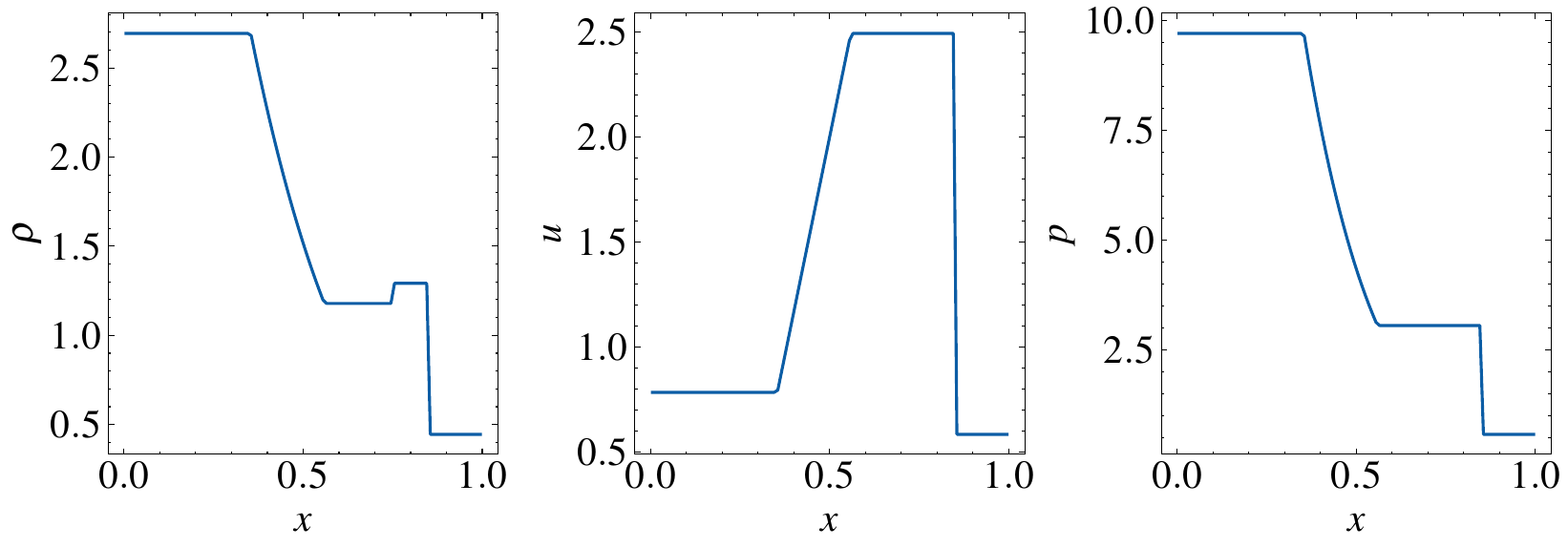}
        \put(-3,36){(b)}
    \end{overpic} \\
    \begin{overpic}[width=0.48\textwidth]{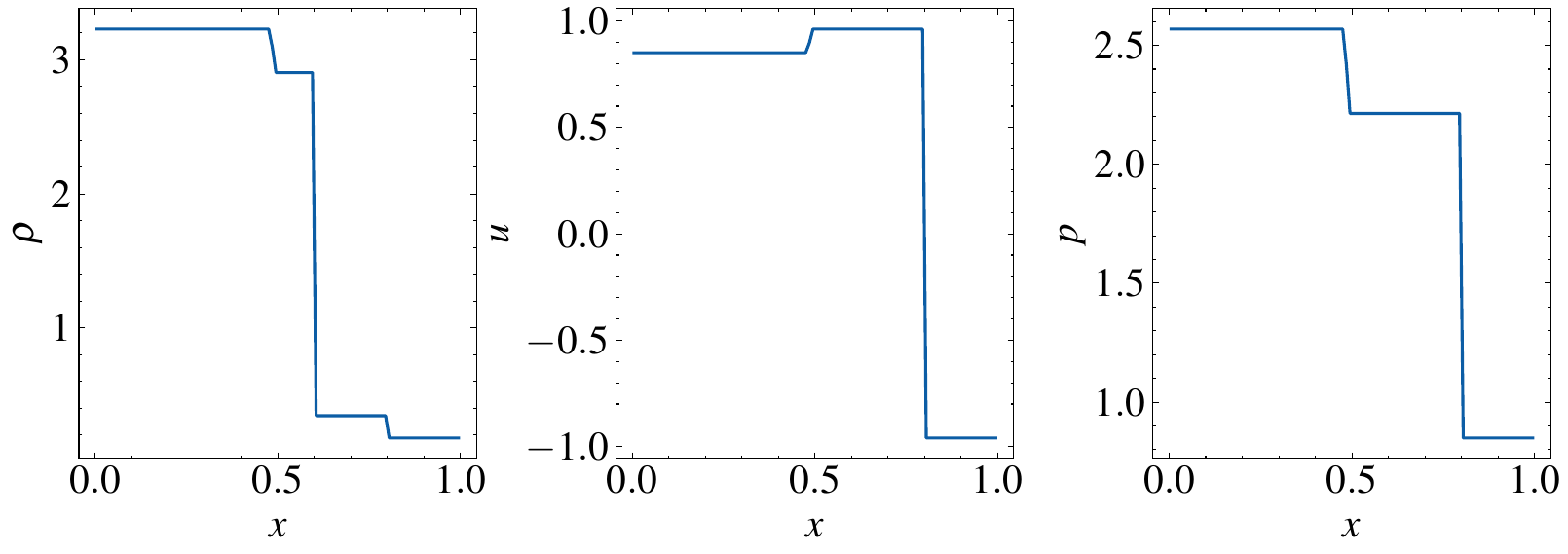}
        \put(-3,36){(c)}
    \end{overpic} &
    \begin{overpic}[width=0.48\textwidth]{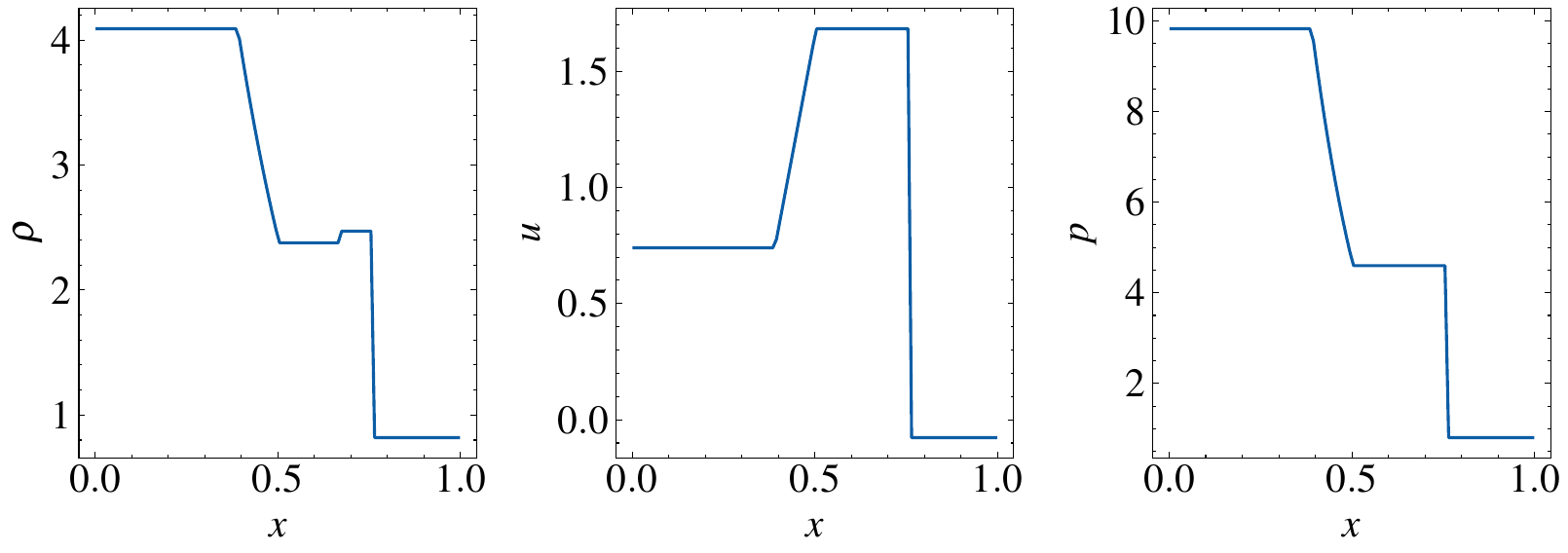}
        \put(-3,36){(d)}
    \end{overpic}
    \end{tabular}
    
    \caption{Visualization of the final density, velocity, and pressure profiles of the training dataset on Sod's shock tube problem with different initial states. All of the solutions in the dataset contain three different characteristics including the rarefaction wave, the contact discontinuity, and the shock discontinuity.}
    \label{fig:euler-dataset}
\end{figure*}

The representative problem for the Euler equations is Sod’s shock tube problem, whose time evolution has an analytical solution producing three characteristic waves: a rarefaction wave, a contact discontinuity, and a shock discontinuity. We randomly generate $20$ initial conditions from uniform distributions within a domain of length $L_x = 1$. The initial condition is defined by left and right primitive states, $Q_l = (\rho_l, u_l, p_l)$ for $x < 0.5$ and $Q_r = (\rho_r, u_r, p_r)$ for $x \geqslant 0.5$. We first sample these states from the following distributions: $\rho_l \sim \mathcal{U}[0.9, 5]$, $\rho_l \sim \mathcal{U}[0.1, 1]$, $u_l, u_r \sim \mathcal{U}[-1, 1]$, $p_l \sim \mathcal{U}[2, 10]$, $p_r \sim \mathcal{U}[0.1, 1]$. Next, we compute the analytical solution at $t=0.1$ s. Finally, we use a test function to screen out any initial conditions whose solution does not exhibit all three characteristic waves, thereby ensuring similarity among the initial conditions in our dataset. Fig.~\ref{fig:euler-dataset} exhibits several sample solutions from the dataset.

\subsection{Training}
As illustrated in Sec.~\ref{sec:diff-sim}, the objective is to find a second-order TVD flux limiter that minimizes the difference between numerical predictions $\hat{Q}$ and true solutions $Q$. Here, we introduce the concept of coarse-graining. While a low-order scheme can achieve high accuracy given sufficiently fine discretization, this approach demands far greater computational resources, making it impractical in many cases. Instead, we aim to employ high-order schemes on a relatively coarse grid, necessitating a flux limiter that performs reliably under such conditions. Therefore, we begin by downsampling the high-resolution initial conditions, originally defined on very fine meshes, to generate their corresponding low-resolution counterparts. We then run simulations with these initial conditions and compare the resulting predictions to the downsampled high-accuracy true solutions. The loss function is defined as the mean squared error (MSE) over the training data:
\begin{eqnarray}
    \mathcal{L} = \frac{1}{STN}\sum_{s=1}^{S}\sum_{n=1}^{T}\sum_{i=1}^{N}\lVert\hat{Q}_{i}^{n,s}-Q_{i}^{n,s}\rVert^2,
\end{eqnarray}
where $S$ denotes the number of trajectories, $T$ denotes the length of each trajectories, and $N$ denotes the number of cells. $\hat{Q}_{i}^{n,s}$ is the predicted solution of cell $i$ at time $n$ for trajectory $s$. For linear advection and Burgers' equation, we downsample the initial conditions with $8\times$ coarse-graining, resulting a resolution of $128$ cells. For Sod's shock tube problem, we discretize the domain into $N = 100 $ cells. 

In light of the strict physical constraints imposed by the simulation, we may consider only the final time snapshot in the loss function, which is equivalent to assigning a weight of $0$ to all earlier snapshots and $1$ to the last snapshot. Doing this can effectively decrease the computational cost during backpropagation. Our experiments also show that the result is not sensitive to the number of snapshots considered in the loss function.

For all three cases, the neural networks consist of $5$ hidden layers, each containing $64$ neurons. $\texttt{ReLU}$ is used as the activation function for the linear advection case, whereas $\texttt{tanh}$ is employed for Burgers’ equation and the Euler equations, for better differentiability considering the nonlinearity of these two problems. We train these networks with the Adam optimizer \citep{kingma2014adam} at a learning rate of $1 \times 10^{-3}$ for $50$ epochs. During this period, the loss converges, and the limiter functions exhibit no notable changes.

\section{Results and discussion}
\subsection{Linear advection}\label{sec:linear-advection}
\begin{figure}[htp]
    \centering
    \begin{overpic}[width=0.7\linewidth]{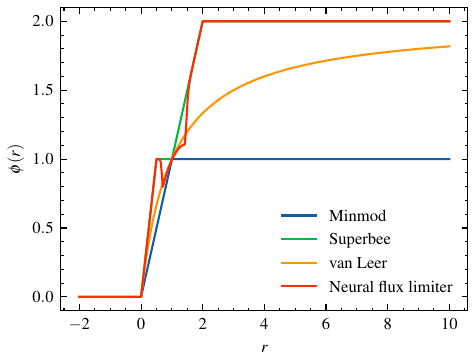}
    \put(0.1,70){(a)}
    \end{overpic}
    \begin{overpic}[width=0.7\linewidth]{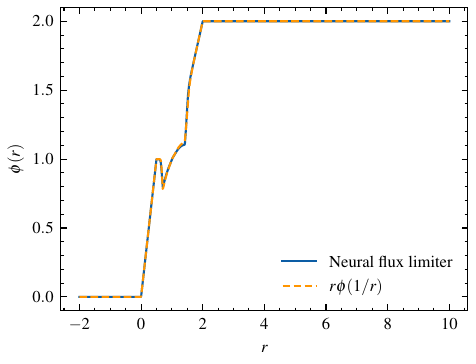}
    \put(0.1,70){(b)}
    \end{overpic}
    
    \caption{The learned flux limiter for the linear advection problem trained on the dataset from PDEBench \citep{takamoto2022pdebench}. (a) Comparison of the neural flux limiter with several classical flux limiters: Minmod, van Leer, and Superbee. Its position within the second-order TVD region demonstrates that the neural flux limiter is both second-order accurate and TVD-compliant. (b) The neural flux limiter function $\phi(r)$ shows perfect agreement with the curve $r\phi(1/r)$, indicating that the symmetry property $\phi(r)/r = \phi(1/r)$ is satisfied. This suggests that the neural network can learn symmetry purely from the data.}
    \label{fig:fl-linear-adv}
\end{figure}

Fig.~\ref{fig:fl-linear-adv}(a) shows the learned neural flux limiter trained on the linear advection dataset, together with other well-known flux limiters. The learned neural flux limiter aligns well with Superbee for values of $r$ away from 1, but differs significantly from Superbee and other classical flux limiters near $r=1$. This behavior suggests that the learned flux limiter minimizes numerical dissipation in non-smooth regions by maximizing the antidiffusive flux, while modifying the curve shape near $(1,1)$ to reduce compressive effects in smooth regions. In addition, the slope of the neural flux limiter at $r=1$ calculated using AD is exactly $0.5$, which coincides with the tangent of van Leer and MC limiters.

Surprisingly, the neural network automatically learned a flux limiter that perfectly satisfies the symmetry property in Eq.~(\ref{eqn:sym-property}), purely from data, as shown in Fig.~\ref{fig:fl-linear-adv}(b). This demonstrates that a flux limiter obeying the symmetry property is more likely to give a better performance.

\begin{figure}[htp]
    \centering
    \begin{overpic}[width=0.75\linewidth]{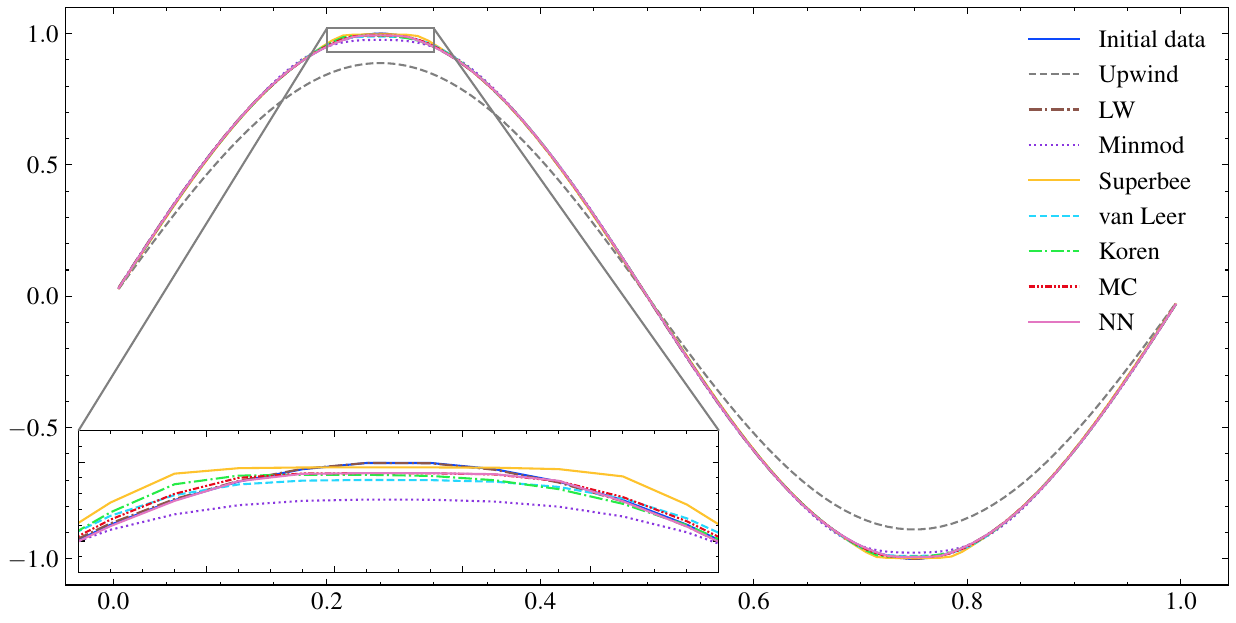}
    \put(-3,45){(a)}
    \end{overpic}
    \begin{overpic}[width=0.75\linewidth]{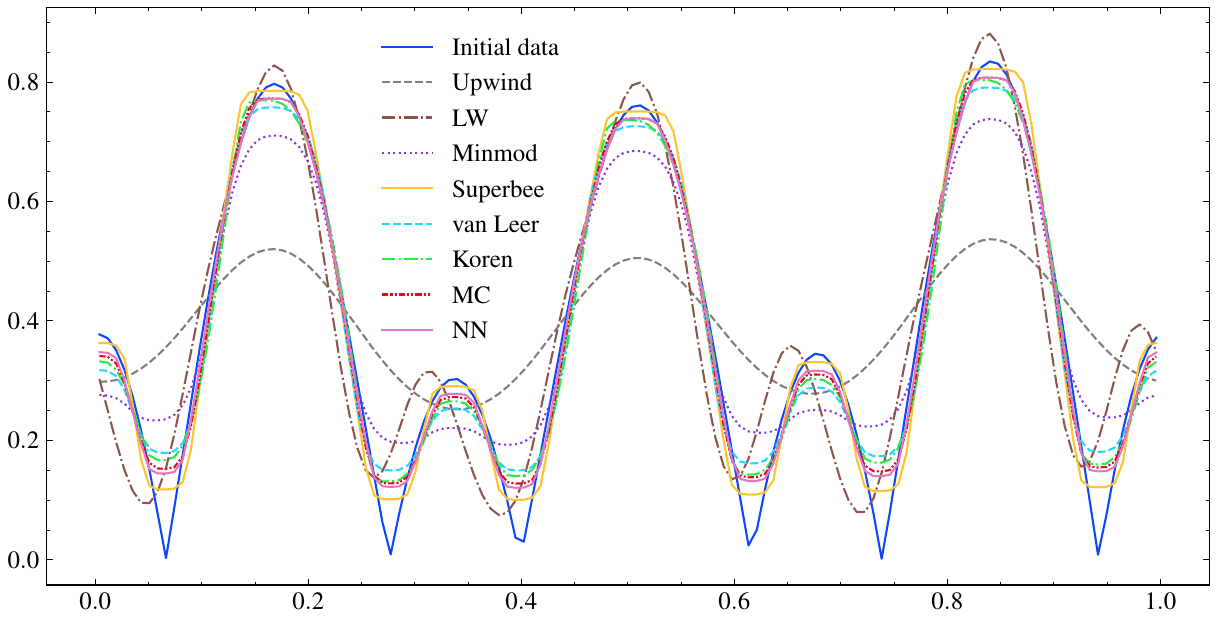}
    \put(-3,45){(b)}
    \end{overpic}
    \begin{overpic}[width=0.75\linewidth]{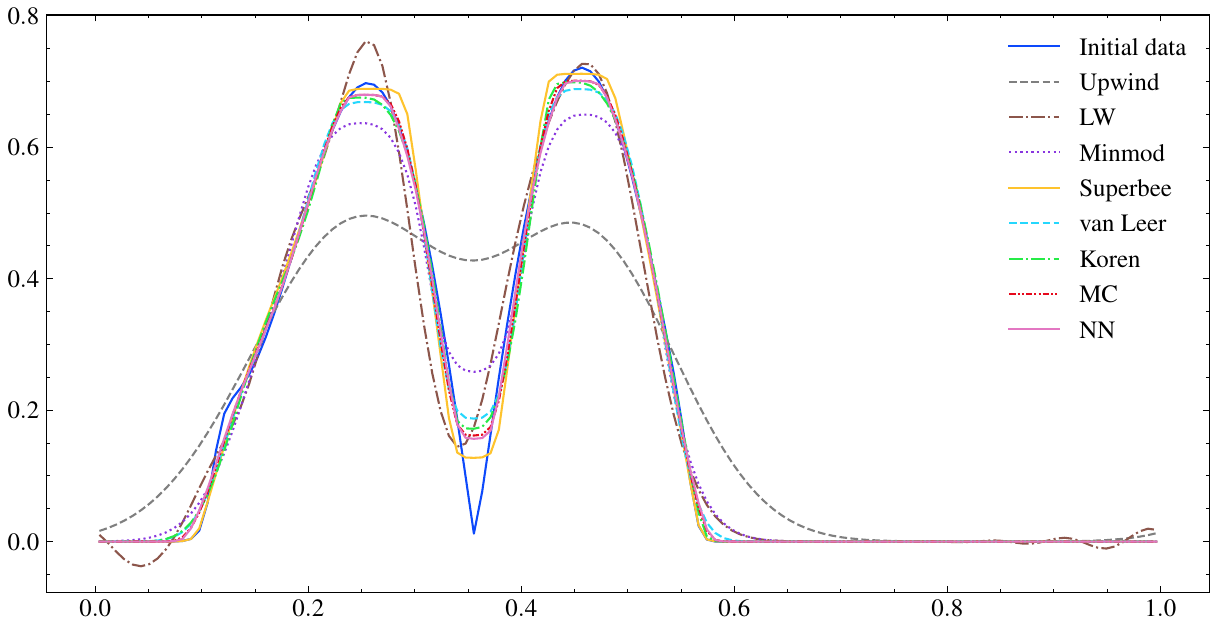}
    \put(-3,45){(c)}
    \end{overpic}
    
    \caption{The performance of the neural flux limiter on linear advection equation with in-distribution initial conditions compared to other classical flux limiters. The solutions are obtained by advecting the initial conditions over one time period with 128 cells. (a) A single sinusoidal wave. (b) A superposition of two sinusoidal waves with absolute value function applied. (c) A superposition of two sinusoidal waves with absolute value function and window function applied.}
    \label{fig:in-distribution-performance-linear-adv}
\end{figure}

\begin{figure}[htp]
    \centering
    \begin{overpic}[width=0.75\linewidth]{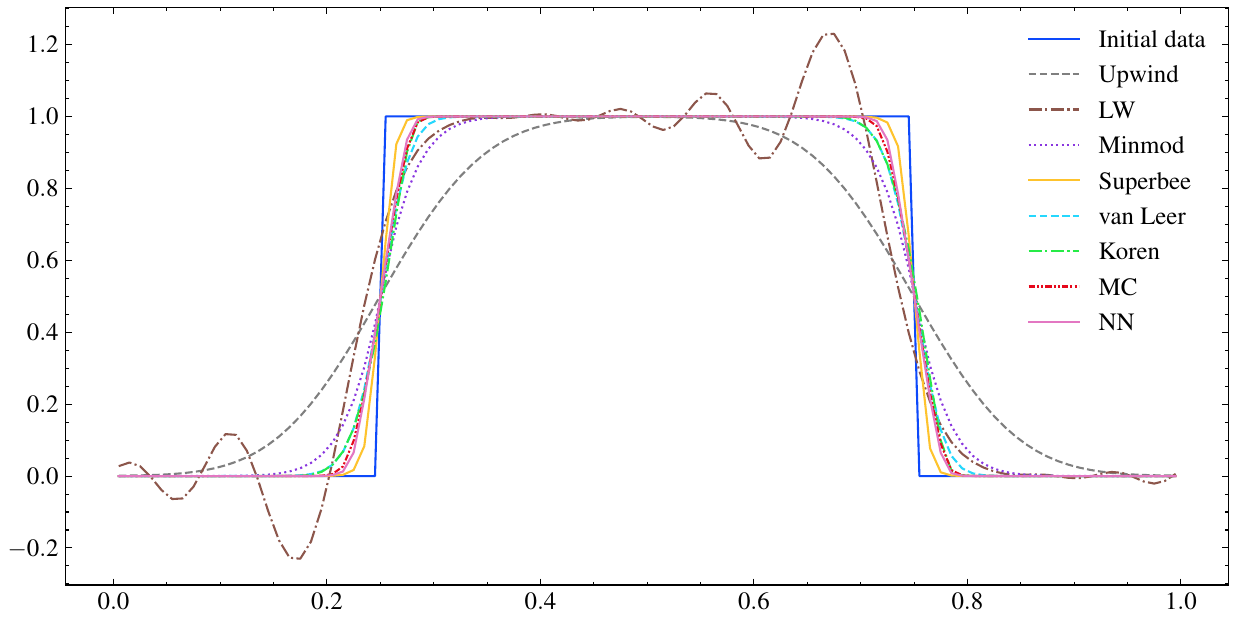}
    \put(-3,45){(a)}
    \end{overpic}
    \begin{overpic}[width=0.75\linewidth]{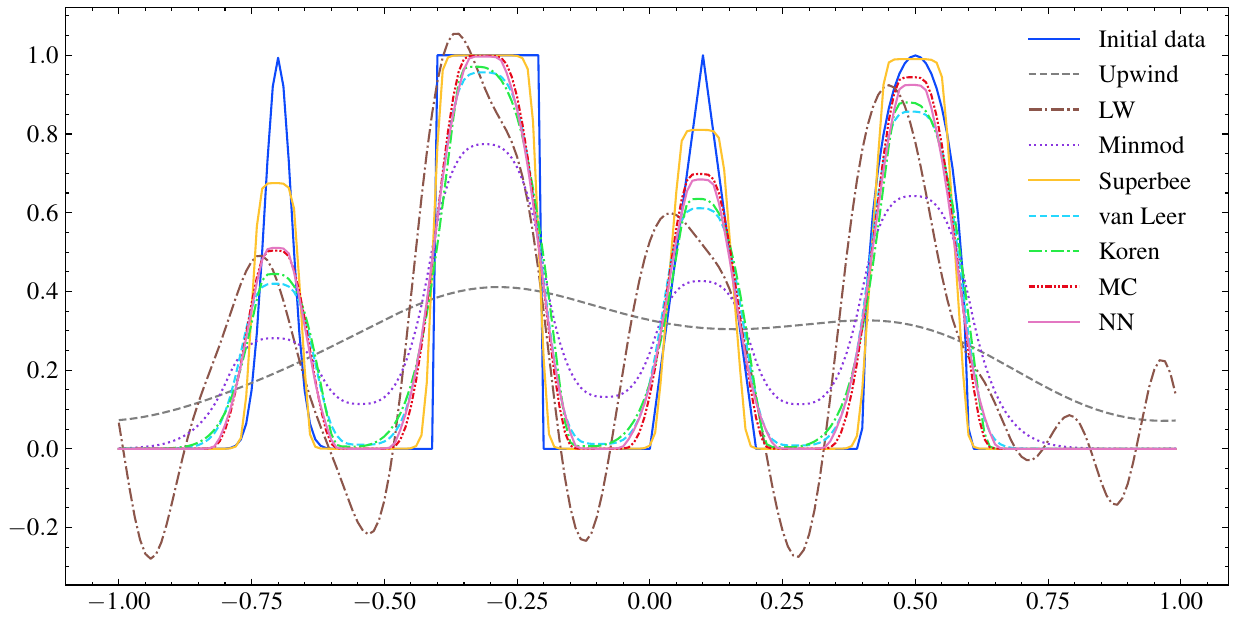}
    \put(-3,45){(b)}
    \end{overpic}
    
    \caption{The performance of the neural flux limiter on linear advection equation with out-of-distribution initial conditions compared to other standard flux limiters. (a) A squara wave with 100 cells is advected over one time period. (b) A combination of a Gaussian, a square wave, a sharp triangle wave, and a half ellipse with 200 cells is advected over 4 time periods ($t=8$ s).}
    \label{fig:out-of-distribution-performance-linear-adv}
\end{figure}

We compare the solution profiles of the final states with the initial states (see Fig.~\ref{fig:in-distribution-performance-linear-adv} for in-distribution initial conditions and Fig.~\ref{fig:out-of-distribution-performance-linear-adv} for out-of-distribution initial conditions) and the mean squared error (see Tab.~\ref{tab:mse-linear-adv}) over the solutions for several representative initial conditions using different flux limiters. Notably, we test the performance over an advection time of at least one full period, which is significantly longer than the one-eighth period used in the training dataset. The mathematical expressions for these standard flux limiters are listed in Tab.~\ref{tab:limiter-expression}
\begin{table}[htp]
\caption{Mathematical expressions for the standard flux limiters used for comparison with the neural flux limiter.}
\centering
\begin{ruledtabular}
\begin{tabular}{cc}
Scheme                               & Expression \\ 
\midrule
Upwind                               & $\phi(r) \equiv 0$ \\
LW                                   & $\phi(r) \equiv 1$ \\
Minmod                               & $\phi(r) = \max(0, \min(1,r))$ \\
Superbee                             & $\phi(r) = \max(0, \min(2r,1), \min(r,2))$ \\
van Leer                             & $\phi(r) = \frac{r+|r|}{1+|r|}$ \\
Koren                                & $\phi(r) = \max(0, \min(2r, \min(\frac{1+2r}{3},2)))$ \\
MC                                   & $\phi(r) = \max(0, \min(2r, \frac{1+r}{2},2))$ \\
\end{tabular}
\end{ruledtabular}
\label{tab:limiter-expression}
\end{table}

For in-distribution initial conditions, our neural flux limiter achieved the best performance over all flux limiters. As shown in Fig.~\ref{fig:in-distribution-performance-linear-adv}, the local extrema in the solution profiles are resolved well. Among second-order TVD limiters excluding Superbee, the peak values of the solutions using the learned flux limiter exhibit the smallest deviation from the reference solution. However, Superbee excessively ``squares off'' the profile, causing undue distortion that may degrade accuracy. In contrast, our learned flux limiter strikes a balance between accuracy and moderation in compressiveness.

For out-of-distribution initial conditions, we consider a single square wave and a standard wave combination \citep{jiang1996efficient} to test the generalizability and discontinuity-capturing ability. The initial condition for the wave combination case is given by
\begin{widetext}
\begin{equation}
    q_0(x) =
    \begin{cases}
    \dfrac{1}{6}\left(G(x,\beta,z-\delta) + G(x,\beta,z+\delta) + 4G(x,\beta,z)\right),
    & -0.8 \leqslant x \leqslant -0.6, \\[6pt]
    
    1, 
    & -0.4 \leqslant x \leqslant -0.2, \\[6pt]
    
    1 - \left|10(x - 0.1)\right|, 
    & 0 \leqslant x \leqslant 0.2, \\[6pt]
    
    \dfrac{1}{6}\left(F(x,\alpha,a-\delta) + F(x,\alpha,a+\delta) + 4F(x,\alpha,a)\right), 
    & 0.4 \leqslant x \leqslant 0.6, \\[6pt]
    
    0, 
    & \text{otherwise},
    \end{cases}
\end{equation}
\end{widetext}
where
\begin{equation}
\begin{aligned}
    G(x,\beta,z) &= e^{-\beta(x-z)^2}, \\
    F(x,\alpha,a) &= \sqrt{\max(1-\alpha^2(x-a)^2, 0)}.
\end{aligned}
\end{equation}
The constants are taken as $a=0.5$, $z=-0.7$, $\delta=0.005$, $\alpha=10$, and $\beta=\ln 2/(36\delta^2)$.

Fig.~\ref{fig:out-of-distribution-performance-linear-adv}(a) shows that the solution using the learned flux limiter is the second closest to the reference, ranking just behind Superbee. This suggests its ability to advect discontinuities. Meanwhile, Fig.~\ref{fig:out-of-distribution-performance-linear-adv}(b) demonstrates that the learned flux limiter outperforms Minmod, van Leer, and Koren, and is competitive with MC under a complex initial condition and a very long advection time ($64$ times the training advection time).

Note that Superbee’s behavior near $(1,1)$ is specifically designed to capture linear discontinuities, enabling them to propagate indefinitely long time without numerical diffusion \citep{roe1983asymptotic}. However, this advantage comes at the cost of ``squaring off'' smooth profiles. Consequently, the overall solution accuracy using Superbee can suffer in regions where the flow remains smooth. In contrast, the learned flux limiter achieves a superior balance, preserving solution quality for both smooth and discontinuous parts of the flow. We are justified to claim that the learned flux limiter generalizes effectively to previously unseen initial conditions and is adept at advecting discontinuities.

\begin{table}
\caption{Comparison of MSE for different flux limiters on linear advection samples given in Fig.\ref{fig:in-distribution-performance-linear-adv} and Fig.\ref{fig:out-of-distribution-performance-linear-adv}. In the in-distribution set the initial conditions are similar to those in the training dataset, while in the out-of-distribution set they are significantly different. The best one for each sample is bolded.}
\centering
\begin{ruledtabular}
\begin{tabular}{cccccc}
Scheme & \multicolumn{5}{c}{MSE}                                                                          \\ \cmidrule{2-6} 
                        & \multicolumn{3}{c}{In-distribution}                    & \multicolumn{2}{c}{Out-of-distribution} \\ \cmidrule(lr){2-4}\cmidrule(lr){5-6} 
                        & (a)              & (b)              & (c)              & (a)                & (b)                \\ \midrule
Upwind                  & 6.24E-3          & 3.24E-2          & 1.38E-2          & 3.61E-2            & 1.29E-1            \\
LW                      & 6.03E-6          & 8.53E-3          & 2.02E-3          & 2.27E-2            & 6.51E-2            \\
Minmod                  & 6.32E-5          & 6.67E-3          & 1.78E-3          & 1.42E-2            & 5.00E-2            \\
Superbee                & 2.96E-5          & 1.62E-3          & 5.62E-4          & \textbf{4.87E-3}   & \textbf{6.93E-3}   \\
van Leer                & 1.05E-5          & 2.81E-3          & 6.57E-4          & 1.00E-2            & 2.12E-2            \\
Koren                   & 9.43E-6          & 2.21E-3          & 5.63E-4          & 1.02E-2            & 2.13E-2            \\
MC                      & 3.06E-6          & 1.67E-3          & 4.24E-4          & 8.97E-3            & 1.57E-2            \\
NN                      & \textbf{2.39E-6} & \textbf{1.32E-3} & \textbf{3.48E-4} & 8.43E-3            & 1.75E-2            \\ 
\end{tabular}
\end{ruledtabular}
\label{tab:mse-linear-adv}
\end{table}

\subsection{Burgers' equation}
In this section, we apply the proposed framework to Burgers' equation. The learned flux limiter smoothly converges to Superbee within $10$ epochs, which indicates that the optimal flux limiter for Burgers' equation is Superbee. This result aligns with the known effectiveness of Superbee in capturing shocks, as elaborated in Sec.~\ref{sec:linear-advection}.

\begin{figure}[htbp]
    \centering
    \includegraphics[width=0.7\linewidth]{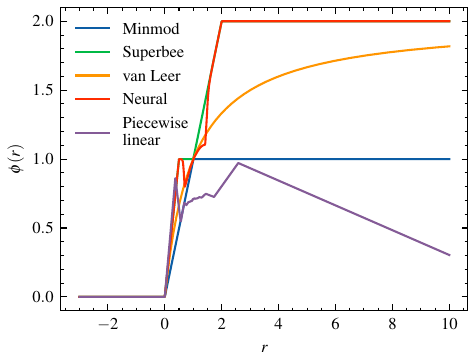}
    \caption{Comparison of our neural flux limiter trained on linear advection equation with the piecewise linear flux limiter trained on Burgers' equation by \citep{nguyen2022machine} using the least square regression over randomly generated dataset, while the optimal flux limiter for Burgers' equation trained on the dataset from \citep{takamoto2022pdebench} by our framework is Superbee. The piecewise linear flux limiter shown in this figure is trained on $2\times$ coarse-grained dataset. It roughly lies in the TVD region but does not pass through the point (1, 1), indicating that the piecewise linear flux limiter is not second-order accurate. This conclusion applies to the other piecewise linear flux limiters with different coarse-grainning.}
    \label{fig:burgers-fl-comparison}
\end{figure}

On the other hand, \citet{nguyen2022machine} also learned an optimal piecewise linear flux limiter for solving Burgers' equation. The authors discretized the limiter function $\phi(r)$ as a piecewise linear function with $K$ segments, where the length and slope of each segment are learnable. To optimize these parameters, they define an MSE cost function that measures the discrepancy between the numerical solution using the parameterized flux limiter and high-resolution reference data. The optimization problem is then reduced to solving an overdetermined linear system of equations, which can be handled using least square regression. The piecewise flux limiter shown in Fig.~\ref{fig:burgers-fl-comparison} is trained on $2\times$ coarse-grained dataset with $K=20$. Clearly, this flux limiter is roughly TVD but not second-order accurate.

In the linear advection problem, we learned the optimal second-order TVD flux limiter for a given dataset with initial conditions satisfying some distribution. In the following discussion, we would also like to assess its ability to generalize to nonlinear scalar conservation laws.

Tab.~\ref{tab:mse-burgers} compares the performance of various flux limiters on Burgers' equation with $1024$ different initial conditions. Among all schemes, Superbee achieves the smallest MSE, confirming our optimization result that Superbee is the optimal flux limiter for Burgers' equation. Notably, our neural flux limiter trained solely on the simpler linear advection equation attains an MSE (7.47E-4) close to that of Superbee (7.35E-4) and outperforms other classic limiters such as Minmod, van Leer, Koren, and MC. This highlights the versatility of our NN model, which generalizes effectively beyond its original training context. In contrast, the piecewise linear approach proposed in \citet{nguyen2022machine} yields an MSE (1.29E-3) larger than most traditional flux limiters, suggesting that enforing $\phi(r)$ to pass through $(1,1)$ is cruicial for obtaining second-order accuracy.

\begin{table}
\caption{Comparison of MSE for different flux limiters on Burgers' equation with 1024 initial conditions. The best and the second best ones are in bold and underlined, respectively.}
\centering
\begin{ruledtabular}
\begin{tabular}{cc}
Scheme                               & MSE                         \\ \midrule
Upwind                               & 3.19E-3                     \\
LW                                   & 4.68E-3                     \\
Minmod                               & 1.06E-3                     \\
Superbee                             & \textbf{7.35E-4}                     \\
van Leer                             & 8.56E-4                     \\
Koren                                & 7.93E-4                     \\
MC                                   & 7.85E-4                     \\
NN (linear advection)                & \underline{7.47E-4}                     \\
Piecewise linear\citep{nguyen2022machine} & 1.29E-3 \\
\end{tabular}
\end{ruledtabular}
\label{tab:mse-burgers}
\end{table}

\subsection{Euler equations}
Having successfully applied our framework to both linear and nonlinear scalar hyperbolic conservation laws, we now aim to determine the optimal flux limiter for Sod's shock tube problem described by the Euler equations, a hyperbolic system of conservation laws.

By optimizing the overall MSE across all 20 solutions illustrated in Sec.~\ref{sec:euler-dataset}, the optimal flux limiter converges to Superbee again within $40$ epochs. This outcome can also be attributed to Superbee’s strong performance in capturing both contact and shock discontinuities, which predominate in the solutions.

\begin{figure}[htbp]
    \centering
    \includegraphics[width=0.7\linewidth]{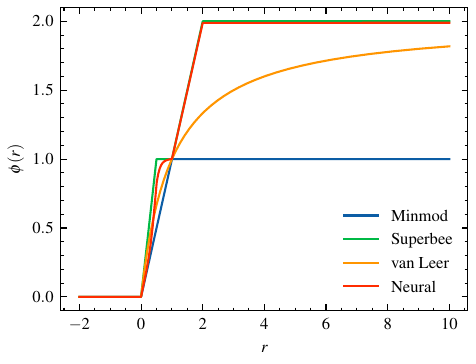}
    \caption{The learned flux limiter for the Euler equations trained on a single trajectory of Sod's shock tube problem, together with several classical flux limiters: Minmod, van Leer, and Superbee.}
    \label{fig:end-to-end-euler}
\end{figure}

However, the learned flux limiter would be different if we perform end-to-end optimization over a single initial condition. For this purpose, we choose the most canonical initial states:
\begin{equation}
(\rho, u, p) =
\begin{cases}
(1.000, 0, 1.0), & x \in [0, 0.5),\\
(0.125, 0, 0.1), & x \in [0.5, 1].\\
\end{cases}
\end{equation}
The corresponding optimal flux limiter is shown in Fig.~\ref{fig:end-to-end-euler}, which exhibits a subtle deviation from Superbee for $0<r<1$. Tab.~\ref{tab:mse-euler-sod} compares the MSEs for different primitive variables using different flux limiters. Despite only training up to $t=0.1$ s, the MSEs using this flux limiter at $t=0.2$ s for all three primitive variables are the lowest among all tested schemes, surpassing even Superbee. One possible explanation is that a single initial condition allows for more targeted optimization, whereas optimization over multiple initial conditions inherently balances performance across varying scenarios.

Additionally, from Tab.~\ref{tab:mse-euler-sod} we also see that the neural flux limiter initially trained on the linear advection problem outperforms every other limiter except Superbee on this problem, suggesting that it can also generalize effectively to hyperbolic systems of conservation laws. Furthermore, Tab.~\ref{tab:mse-euler-lax-shu-osher} compares the MSEs evaluated on two benchmark problems, Lax's problem
\begin{equation}
(\rho, u, p) =
\begin{cases}
(0.445, 0.698, 3.528), & x \in [0, 0.5),\\
(0.500, 0.000, 0.571), & x \in [0.5, 1],\\
\end{cases}
\end{equation}
and the Shu--Osher's problem
\begin{equation}
(\rho, u, p) =
\begin{cases}
(3.857143, 2.629369, 10.33333), & x \in [-5, -4),\\
(1+0.2\sin(5x), 0, 1), & x \in [-4, 5].\\
\end{cases}
\end{equation}
Though superbee is the optimal flux limiter for these two problems, our nerual flux limiter trained on linear advection equation is consistently better than others, indicating robust generalization across various test cases.

\begin{table}
\caption{Comparison of MSE for different flux limiters on Sod's shock tube problem. The best one for each primitive variable is bolded.}
\centering
\begin{ruledtabular}
\begin{tabular}{cccc}
Scheme   & \multicolumn{3}{c}{MSE}     \\
\cmidrule{2-4}
         & $\rho$     & $u$       & $p$       \\
\midrule
Upwind   & 6.73E-4 & 4.38E-3 & 7.36E-4 \\
Minmod   & 2.33E-4 & 1.55E-3 & 2.07E-4 \\
Superbee & 1.61E-4 & 1.36E-3 & 1.74E-4 \\
van Leer & 1.95E-4 & 1.40E-3 & 1.88E-4 \\
Koren    & 1.91E-4 & 1.34E-3 & 1.91E-4 \\
MC       & 1.89E-4 & 1.39E-3 & 1.89E-4 \\
NN (linear advection)      & 1.78E-4 & 1.36E-3 & 1.83E-4 \\
NN (end-to-end)      & \textbf{1.59E-4} & \textbf{1.15E-3} & \textbf{1.63E-4} \\
\end{tabular}
\end{ruledtabular}
\label{tab:mse-euler-sod}
\end{table}

\begin{table*}
\caption{Comparison of MSE for different flux limiters on Lax's and the Shu-Osher's problem. The best and the second best ones are in bold and underlined, respectively. Though Superbee is the optimal flux limiter for this problem, our nerual flux limiter trained on linear advection equation is consistently better than others, which indicates that the learned flux limiter has great generalization ability.}
\centering
\begin{ruledtabular}
\begin{tabular}{ccccccc}
Scheme               & \multicolumn{6}{c}{MSE}                                                                                                          \\ \cmidrule{2-7} 
\multicolumn{1}{l}{} & \multicolumn{3}{c}{Lax}                                & \multicolumn{3}{c}{Shu-Osher}                                           \\ \cmidrule(lr){2-4}\cmidrule(lr){5-7}
       & $\rho$     & $u$       & $p$                & $\rho$     & $u$       & $p$ \\ \midrule
Upwind               & 9.79E-3          & 1.55E-2          & 1.92E-2          & 5.56E-2                 & 4.52E-2               & 3.95E-1               \\
Minmod               & 3.56E-3          & 4.42E-3          & 6.82E-3          & 3.17E-2                 & 1.41E-2               & 1.52E-1               \\
Superbee             & \textbf{1.81E-3} & \textbf{2.06E-3} & \textbf{5.29E-3} & \textbf{2.82E-2}        & \textbf{5.30E-3}      & \textbf{1.074E-1}     \\
van Leer             & 2.70E-3          & 3.03E-3          & 5.96E-3          & 2.93E-2                 & 8.70E-3               & 1.25E-1               \\
Koren                & 2.56E-3          & 2.65E-3          & 5.77E-3          & 2.90E-2                 & 7.08E-3               & 1.18E-1               \\
MC                   & 2.50E-3          & 2.55E-3          & 5.74E-3          & 2.89E-2                 & 6.83E-3               & 1.17E-1               \\
NN (linear advection)                   & \underline{2.31E-3}          & \underline{2.26E-3}          & \underline{5.54E-3}          & \underline{2.84E-2}                 & \underline{5.62E-3}               & \underline{1.11E-1}               \\
\end{tabular}
\end{ruledtabular}
\label{tab:mse-euler-lax-shu-osher}
\end{table*}

\subsection{Two-dimensional Riemann problem}
We have seen that the neural flux limiter trained solely on the linear advection problem can be well applied to one-dimensional nonlinear scalar hyperbolic conservation laws and hyperbolic systems of conservation laws. We now want to assess the generalizability on two-dimensional problems, particularly the two-dimensional Riemann problem for gas dynamics. In this section, we focus on a classical setup in a square domain where the initial condition is divided into four quadrants, each initialized with distinct constant states. This serves as a stringent test of the flux limiter’s ability to maintain stability and capture multi-dimensional wave interactions.

\begin{figure}[htp]
    \centering
    \begin{minipage}[c]{0.49\linewidth}
        \begin{overpic}[width=\textwidth]{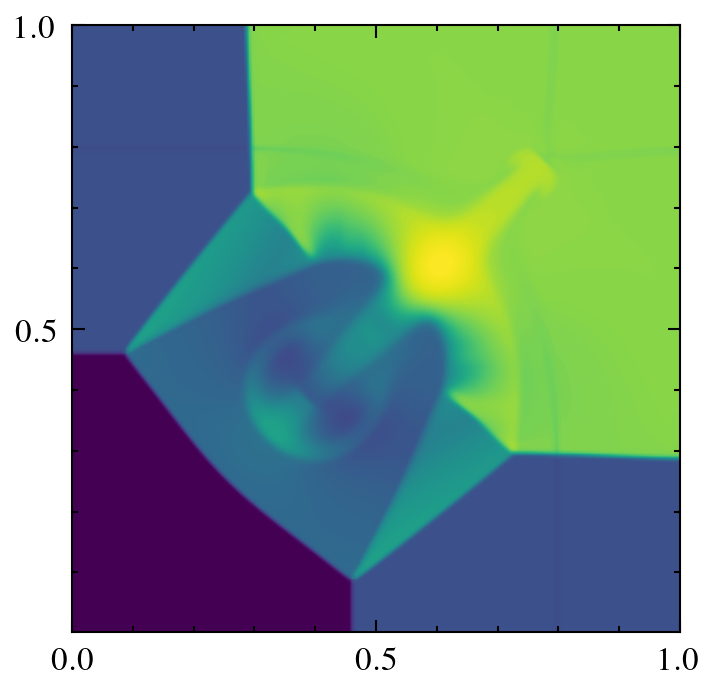}
        \put(2,85){(a)}
        \end{overpic}
    \end{minipage}
    \begin{minipage}[c]{0.49\linewidth}
        \begin{overpic}[width=\textwidth]{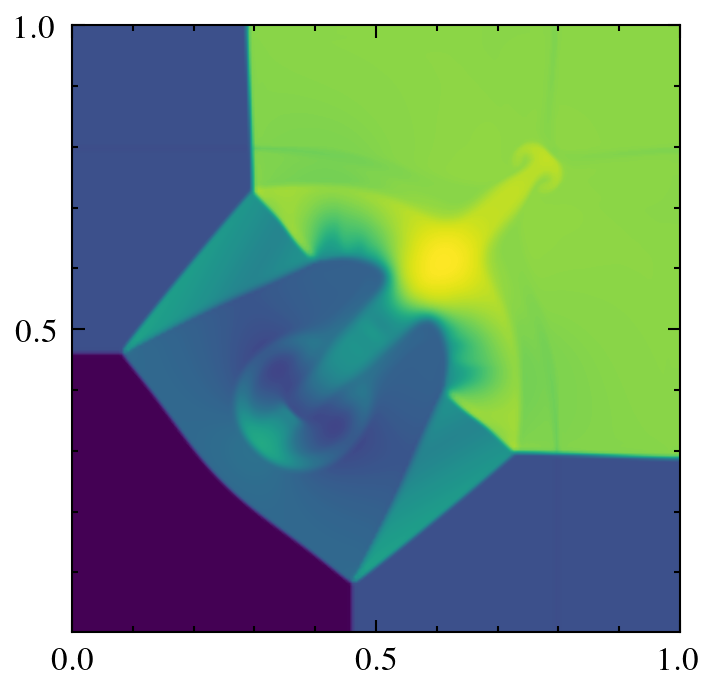}
        \put(2,85){(b)}
        \end{overpic}
    \end{minipage}
    \begin{minipage}[c]{0.49\linewidth}
        \begin{overpic}[width=\textwidth]{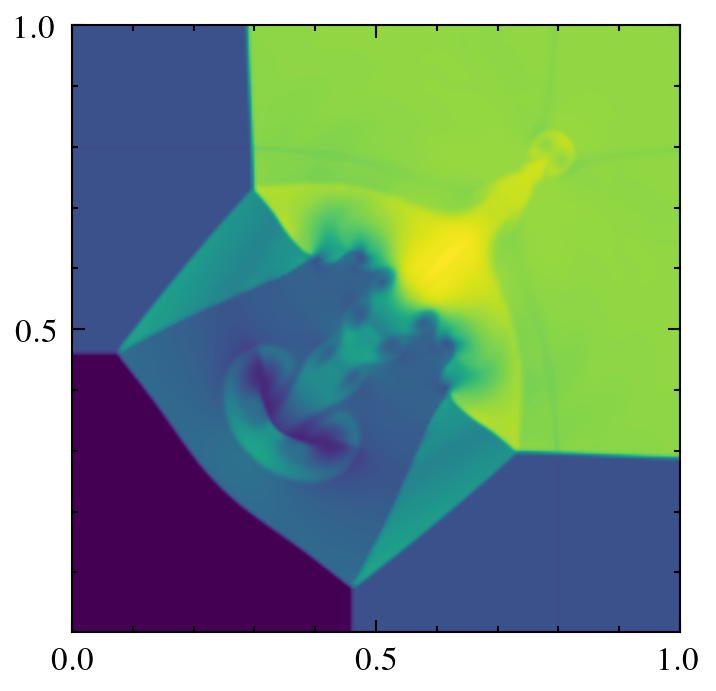}
        \put(2,85){(c)}
        \end{overpic}
    \end{minipage}
    \begin{minipage}[c]{0.49\linewidth}
        \begin{overpic}[width=\textwidth]{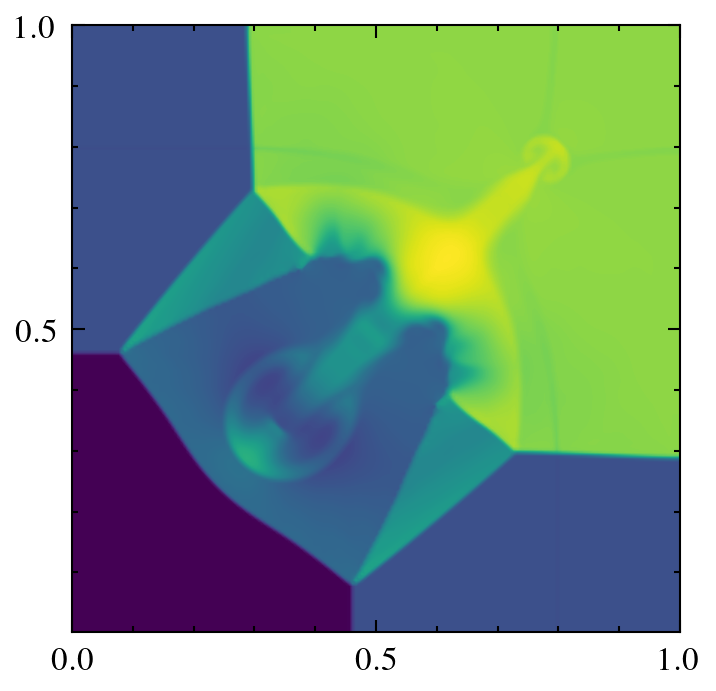}
        \put(2,85){(d)}
        \end{overpic}
    \end{minipage}
    
    \caption{Comparison of the density contours at $t=0.8$ s of the two-dimensional Riemann problem obtained using different limiters. The number of cells in $x$ and $y$ direction is 200. (a) Minmod. (b) van Leer. (c) Superbee. (d) Neural flux limiter trained on linear advection equation.}
    \label{fig:2d-riemann-density}
\end{figure}

The two-dimensional Euler equations are given by
\begin{equation}
    \frac{\partial}{\partial t}
    \begin{bmatrix}
        \rho \\
        \rho u \\
        \rho v \\
        E
    \end{bmatrix}
    +
    \frac{\partial}{\partial x}
    \begin{bmatrix}
        \rho u \\
        \rho u^2 + p \\
        \rho uv \\
        (E+p)u
    \end{bmatrix}
    +
    \frac{\partial}{\partial y}
    \begin{bmatrix}
        \rho v \\
        \rho vu \\
        \rho v^2 + p \\
        (E+p)v
    \end{bmatrix}
    = 0.
\end{equation}
The initial conditions are as follows:
\begin{equation}
(\rho, u, v, p) =
\begin{cases}
(1.500, 0.000, 0.000, 1.500), & 0.8 \leqslant x \leqslant 1, 0.8 \leqslant y \leqslant 1,\\
(0.532, 1.206, 0.000, 0.300), & 0 \leqslant x \leqslant 0.8, 0.8 \leqslant y \leqslant 1,\\
(0.138, 1.206, 1.206, 0.029), & 0 \leqslant x \leqslant 0.8, 0 \leqslant y \leqslant 0.8,\\
(0.532, 0.000, 1.206, 0.300), & 0.8 \leqslant x \leqslant 1, 0 \leqslant y \leqslant 0.8.\\
\end{cases}
\end{equation}
Figure~\ref{fig:2d-riemann-density} provides a qualitative comparison of the density fields at $t=0.8$ s for the two-dimensional Riemann problem solved by the wave-propagation algorithm for multidimensional systems documented in \cite{leveque2002finite}. Overall, the neural flux limiter, trained solely on the one-dimensional linear advection equation, successfully captures the main flow structures, including shocks, contact surfaces, and rarefactions, and offers sharper resolution of fine features compared to Minmod and van Leer. While Superbee appears to offer greater detail in the density field, such an advantage may result from its intrinsic over-compressiveness---a characteristic that may not always align with physical reality. Despite the increased complexity of multi-dimensional wave interactions, the neural flux limiter remains robust: discontinuities are well-resolved, and spurious oscillations are kept under control. This result underscores the ability of the neural flux limiter to generalize beyond the very simple one-dimensional linear advection for which it was initially trained, suggesting that it encodes sufficient information to handle the higher-dimensional dynamics present in gas dynamics problems.

\section{Conclusions}
In this work, we develop a data-driven methodology to learn optimal second-order TVD flux limiters via fully differentiable simulations. By representing the flux limiter as a pointwise convex combination of the Minmod and Superbee limiters, we strictly enforce both the second-order accuracy requirement and the TVD property at all stages of training. This ensures that the learned limiter is guaranteed to remain within the Sweby region, avoiding undesirable oscillations while retaining high-resolution capabilities.

For the one-dimensional linear advection problem, our learned flux limiter outperforms all standard second-order TVD limiters, demonstrating excellent balance between accuracy in smooth regions and discontinuity-capturing capability. For the Burgers’ equation and the standard Sod’s shock tube problem of the Euler equations, our approach identifies Superbee as the best limiter to capture shocks. Yet, by performing end-to-end training on a specific Euler problem instance, we discover a learned flux limiter that slightly outperforms even Superbee. These results underscore the versatility and robustness of the proposed framework: the same strategy readily pinpoints an optimal limiter for each problem setup.

A key observation from our experiments is that the flux limiter trained only on the simplest case---the one-dimensional linear advection equation---exhibits remarkable generalizability. Despite being exposed to seemingly simple dynamics, the training data contains a wide diversity of slope ratios, $r$, which paves the road for the limiter’s broader applicability. Indeed, the learned limiter demonstrates favorable performance on a variety of more complex hyperbolic problems, including the Burgers’ equation, Sod’s shock-tube problem, and even two-dimensional Riemann problems. In many cases, our approach outperforms or closely matches well-established flux limiters such as Minmod, van Leer, Koren, and MC, while maintaining competitive shock-capturing ability comparable to Superbee.

Moreover, by performing end-to-end optimization on a single trajectory for Sod’s shock tube, we can train problem-specific limiters that surpass well-known classical limiters, underlining the potential of our approach when fine-tuned for particular applications. For practitioners in the CFD community, our framework makes it feasible to verify whether the previously chosen flux limiter is truly optimal and to select the best option for each specific problem configuration.

Finally, we have integrated the learned limiters into OpenFOAM (see Appendix \ref{app:openfoam}) and can similarly integrate them into other codes.  The only change involves replacing the limiter function with the learned function, which remains computationally lightweight. As modern hardware and AD frameworks mature, we anticipate that such end-to-end differentiable strategies will become increasingly accessible, enabling practitioners to quickly design, train, and validate specialized flux limiters tailored to diverse fluid flow problems.

\begin{acknowledgments}
The authors acknowledge support from Los Alamos National Laboratory. This research was supported in part through computational resources and services provided by Advanced
Research Computing at the University of Michigan, Ann Arbor. This work used Bridges-2 at Pittsburgh Supercomputing Center through allocation CTS180061 from the Advanced Cyberinfrastructure Coordination Ecosystem: Services \& Support (ACCESS) program, which is supported by U.S. National Science Foundation grants \#2138259, \#2138286, \#2138307, \#2137603, and \#2138296. This research was supported by grants from NVIDIA.
\end{acknowledgments}

\appendix

\section{FV schemes for Burgers' equation and Euler equations}\label{app:fvm}

\subsection{Burgers' equation}
For the Burgers' equation
\begin{equation}
    \frac{\partial q}{\partial t} + \frac{\partial}{\partial x}\left(\frac{q^2}{2}\right) = 0,
\end{equation}

the underlying first-order scheme is chosen as the Engquist--Osher scheme \citep{engquist1980stable} which has numerical flux
\begin{equation}
    F_{i-1/2}^{\text{EO}} = f_{i-1}^{+} + f_{i}^{-} + f(\bar{q}),
\end{equation}
where
\begin{equation}
    f_{i}^{\pm} = f(q_{i}^{\pm}), \quad q_{i}^{+} = \max (q_{i}, 0), \quad q_{i}^{-} = \min (q_{i}, 0),
\end{equation}
and $\bar{q}$ is the sonic point of $f(q)$, i.e., $f^{\prime}(\bar{q})=0$. Similar to linear advection, we add both limited positive and negative fluxes to the first-order scheme to obtain a high-resolution scheme
\begin{equation}
\begin{aligned}
    F_{i-1/2} = F_{i-1/2}^{\text{EO}} &+ \phi(r_{i-1}^{+})\alpha_{i-1/2}^{+}\left(\Delta f_{i-1/2}\right)^{+} \\
    &- \phi(r_{i}^{-})\alpha_{i-1/2}^{-}\left(\Delta f_{i-1/2}\right)^{-},
\end{aligned}
\end{equation}
where
\begin{equation}\label{eqn:r-burgers}
\begin{aligned}
    \left(\Delta f_{i-1/2}\right)^{\pm} &= f_{i}^{\pm} - f_{i-1}^{\pm}, \\
    \alpha_{i-1/2}^{\pm} &= \frac{1}{2}\left(1 \mp \nu_{i-1/2}^{\pm}\right), \\
    \nu_{i-1/2}^{\pm} &= \frac{\Delta t}{\Delta x}\frac{\left(\Delta f_{i-1/2}\right)^{\pm}}{\Delta Q_{i-1/2}}, \\
    r_{i}^{\pm} &= \frac{\alpha_{i \mp 1/2}^{\pm}\left(\Delta f_{i \mp 1/2}\right)^{\pm}}{\alpha_{i \pm 1/2}^{\pm}\left(\Delta f_{i \pm 1/2}\right)^{\pm}}.
\end{aligned}
\end{equation}

\subsection{Euler equations}
Consider the Euler equations
\begin{equation}
    \frac{\partial}{\partial t}
    \begin{bmatrix}
        \rho \\
        \rho u \\
        E
    \end{bmatrix}
    +
    \frac{\partial}{\partial x}
    \begin{bmatrix}
        \rho u \\
        \rho u^2 + p \\
        (E+p)u
    \end{bmatrix}
    = 0,
\end{equation}
where $\rho, u, p$, and $E$ denote the density, velocity, pressure, and total energy, respectively. In our numerical solver, we implement the wave-propagation algorithm detailed by \cite{leveque2002finite}, which extends Godunov's method and its variants based on approximate Riemann solvers to high-resolution schemes. The solution of the Riemann problem at cell interface $i-1/2$ given the left and right states $Q_{i-1}$ and $Q_{i}$ typically contains a set of $m$ waves $\mathcal{W}_{i-1/2}^p \in \mathbb{R}^m $ propagating at some speeds $s_{i-1/2}^p \in \mathbb{R}$, with
\begin{equation}
    \Delta Q_{i-1/2} \equiv Q_{i} - Q_{i-1} = \sum_{p=1}^{m}\mathcal{W}_{i-1/2}^p,
\end{equation}
where $m$ indicates the number of equations. Using the waves and speeds from the approximate Riemann solution, we define the left- and right-going fluctuations as
\begin{equation}
\begin{aligned}
    \mathcal{A}^{-}\Delta Q_{i-1/2} &= \sum_{p=1}^{m}(s_{i-1/2}^{p})^{-}\mathcal{W}_{i-1/2}^p, \\
    \mathcal{A}^{+}\Delta Q_{i-1/2} &= \sum_{p=1}^{m}(s_{i-1/2}^{p})^{+}\mathcal{W}_{i-1/2}^p, \\
\end{aligned}
\end{equation}
where
\begin{equation}
\begin{aligned}
    (s_{i-1/2}^{p})^{-} &= \min(s_{i-1/2}^{p}, 0), \\
    (s_{i-1/2}^{p})^{+} &= \max(s_{i-1/2}^{p}, 0).
\end{aligned}
\end{equation} 
The condition
\begin{equation}
    \mathcal{A}^{-}\Delta Q_{i-1/2} + \mathcal{A}^{+}\Delta Q_{i-1/2} = f(Q_i) - f(Q_{i-1})
\end{equation}
guarantees that the solution to the Riemann problem is conservative. With these definitions, the update formula of the states takes the form
\begin{equation}\label{eqn:wave-prop-update}
\begin{aligned}
    Q_{i}^{n+1} = Q_{i}^{n} &- \frac{\Delta t}{\Delta x}(\mathcal{A}^{-}\Delta Q_{i+1/2} + \mathcal{A}^{+}\Delta Q_{i-1/2}) \\
    &- \frac{\Delta t}{\Delta x}(\tilde{F}_{i+1/2} - \tilde{F}_{i-1/2}),
\end{aligned}
\end{equation}
where the flux $\tilde{F}_{i-1/2}$ is the high-resolution correction, which is similar to that in the Lax--Wendroff scheme in Eq.~(\ref{eqn:linear-fou-lw}). It is given by
\begin{equation}
    \tilde{F}_{i-1/2} = \frac{1}{2}\sum_{p=1}^{m}|s_{i-1/2}^{p}|\left(1-\frac{\Delta t}{\Delta x}|s_{i-1/2}^{p}|\right)\tilde{\mathcal{W}}_{i-1/2}^{p}.
\end{equation}
Here, $\tilde{\mathcal{W}}_{i-1/2}^{p}$ represents a limited version of the wave $\mathcal{W}_{i-1/2}^{p}$. It is obtained by first projecting the vector $\mathcal{W}_{I-1/2}^{p}$ in the upwind direction onto $\mathcal{W}_{i-1/2}^{p}$ and then comparing the lengths of this projection and $\mathcal{W}_{i-1/2}^{p}$. Mathematically, the limiting process is expressed as
\begin{equation}
    \tilde{\mathcal{W}}_{i-1/2}^{p} = \phi(r_{i-1/2}^{p})\mathcal{W}_{i-1/2}^{p},
\end{equation}
with
\begin{equation}\label{eqn:r-euler}
\begin{aligned}
    r_{i-1/2}^{p} &= \frac{\langle \mathcal{W}_{I-1/2}^{p}, \mathcal{W}_{i-1/2}^{p} \rangle}{\langle \mathcal{W}_{i-1/2}^{p}, \mathcal{W}_{i-1/2}^{p} \rangle},
    \\
    I &= \left\{
    \begin{array}{cc}
        i-1 & \text{if } s_{i-1/2}^{p}>0,\\
        i+1 & \text{if } s_{i-1/2}^{p}<0.
    \end{array}
    \right.
\end{aligned}
\end{equation}

We choose Roe's method as the approximate Riemann solver, and Eq.~(\ref{eqn:wave-prop-update}) can be written equivalently in the form of Eq.~(\ref{eqn:state-update}) by defining
\begin{equation}
    F_{i-1/2} = F_{i-1/2}^{\text{Roe}} + \tilde{F}_{i-1/2},
\end{equation}
where the Roe flux is
\begin{equation}
    F_{i-1/2}^{\text{Roe}} = \frac{1}{2}\left[f(Q_{i-1})+f(Q_{i})\right] - \frac{1}{2}|A_{i-1/2}|(Q_{i}-Q_{i-1}),
\end{equation}
and $A_{i-1/2}$ is the Roe linearization matrix at cell interface $i-1/2$. The notation $|A_{i-1/2}|$ refers to taking the absolute values of the eigenvalues in the eigendecomposition of $A_{i-1/2}$.

\section{Integrate the nerual flux limiter into OpenFOAM}\label{app:openfoam}

\begin{figure}[htbp]
    \centering
    \begin{overpic}[width=0.68\linewidth]{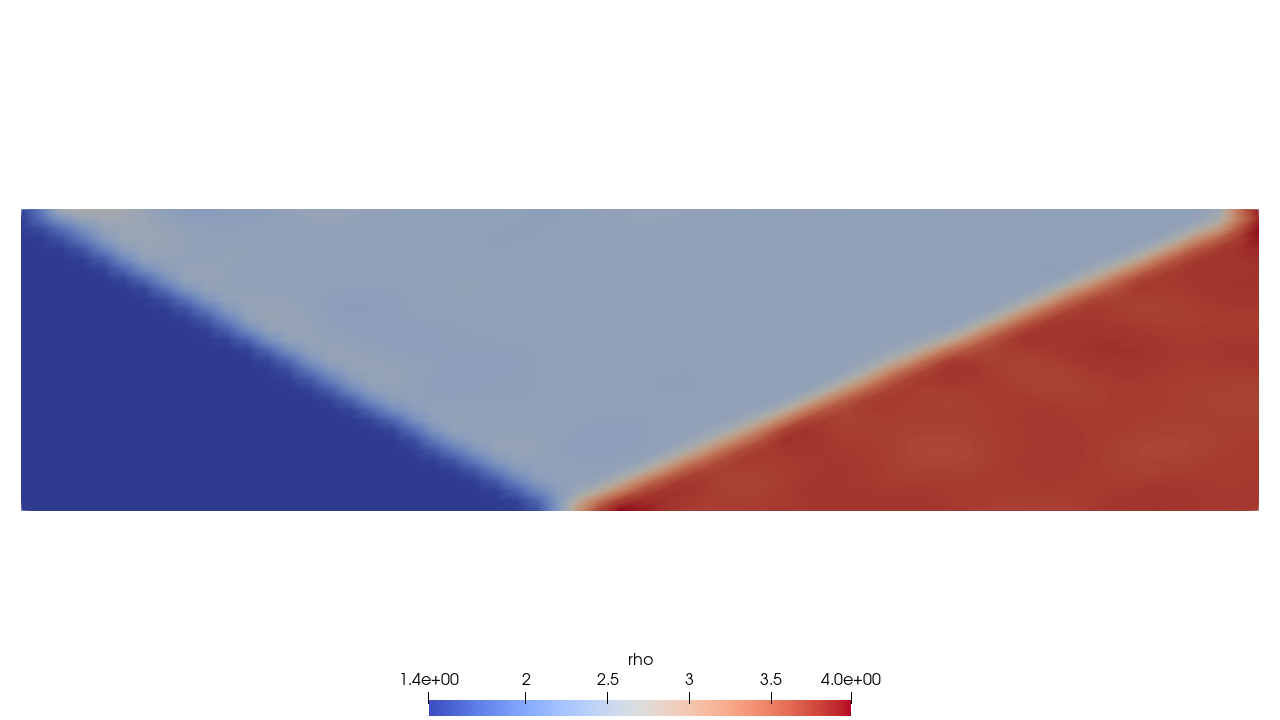}
    \put(-3,45){(a)}
    \end{overpic}
    \begin{overpic}[width=0.68\linewidth]{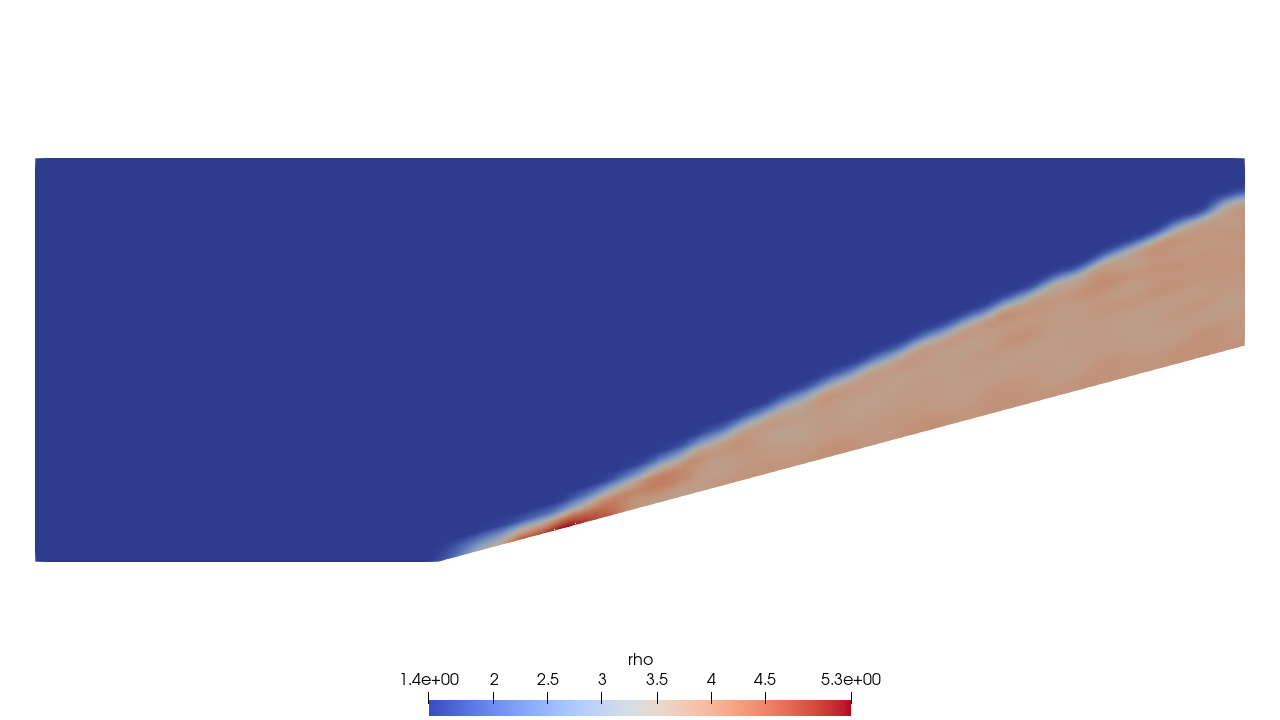}
    \put(-3,45){(b)}
    \end{overpic}
    \begin{overpic}[width=0.68\linewidth]{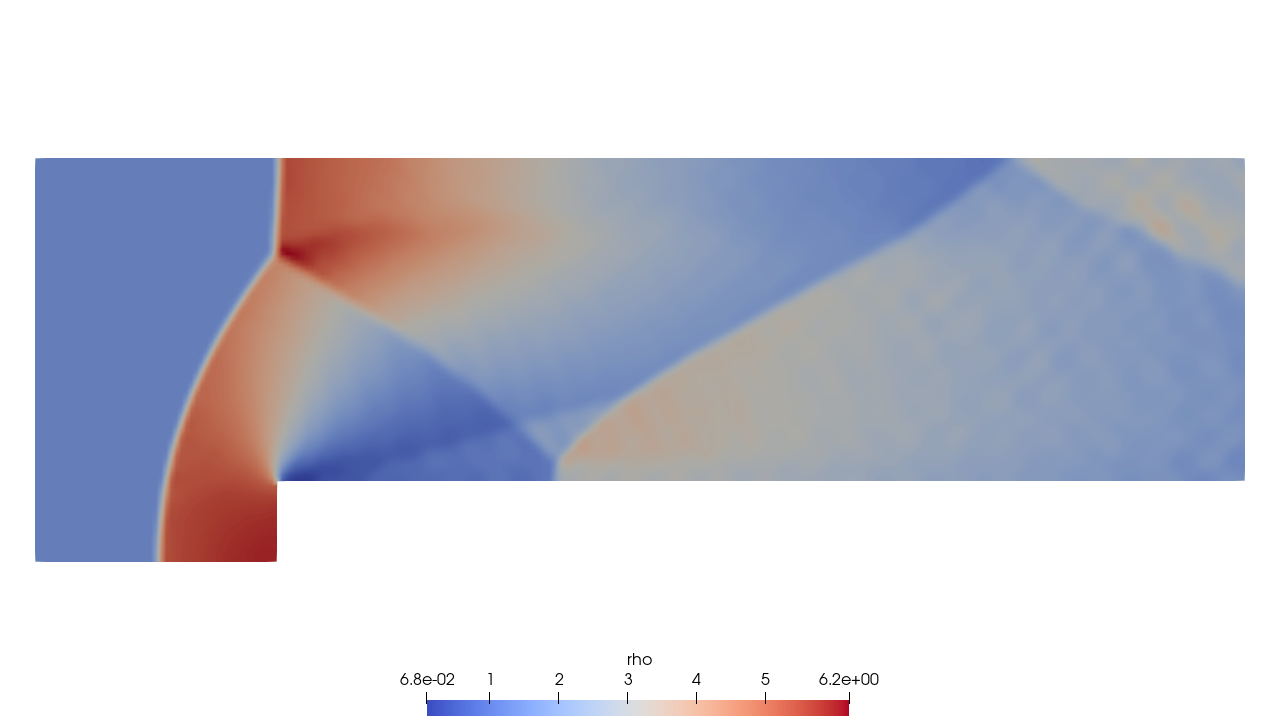}
    \put(-3,45){(c)}
    \end{overpic}
    \caption{The density fields at the end of simulation for three benckmark problems in OpenFOAM. (a) The Mach 2.9 gas flow entering horizontally from the left collides with the slanted Mach 2.66 gas flow from the top, forming an oblique shock. (b) A Mach 5 supersonic flow encounters a 15-degree inclined wedge, producing an oblique shock wave as it deflects around the wedge surface. (c) A Mach 3 flow enters a rectangular geometry with a step near the inlet, generating shock waves.}
    \label{fig:openfoam-sims}
\end{figure}

In this Appendix, we integrate the neural flux limiter trained on linear advection equation into OpenFOAM and show the numerical results of three benchmark problems in OpenFOAM's tutorials. In the first two cases, we apply our neural flux limiter to all three fields: density, velocity, and temperature. For the forward-step case, however, we use van Leer limiter for the density and temperature fields, and employ the neural flux limiter solely for the velocity field. Fig.~\ref{fig:openfoam-sims} shows the density fields at the final simulation time for all three cases. The neural limiter successfully captures the shock waves without inducing spurious oscillations near discontinuities, while maintaining good resolution in smooth regions.

\bibliography{references}

\end{document}